\documentclass[aps,prb,twocolumn,floatfix,superscriptaddress,showpacs]{revtex4}
\usepackage{times}
\usepackage{epsfig}
\usepackage{amsmath}
\usepackage{dsfont}
\usepackage{bbm}
\usepackage{amssymb}
\usepackage[colorlinks=true,citecolor=red,linkcolor=blue]{hyperref}

\newcommand{\bs}{\boldsymbol}

\begin{document}

\title{Sum rule violation in self-consistent hybridization expansions}

\author{Andreas R\"uegg}
\affiliation{Department of Physics, The University of Texas at Austin, Austin, TX 78712, USA}
\affiliation{Department of Physics, University of California, Berkeley, CA 94720, USA}
\author{Emanuel Gull}
\affiliation{Department of Physics, Columbia University, New York, NY 10027, USA}
\affiliation{Max-Plank Institute for Complex Systems, Dresden, Germany}
\affiliation{Department of Physics, University of Michigan, Ann Arbor, MI 48109, USA}
\author{Gregory A.~Fiete}
\affiliation{Department of Physics, The University of Texas at Austin, Austin, TX 78712, USA}
\author{Andrew J.~Millis}
\affiliation{Department of Physics, Columbia University, New York, NY 10027, USA}

\begin{abstract}
We show that for multi-orbital quantum impurity models the non-crossing approximation and one-crossing approximation versions of the  self-consistent hybridization expansions violate the sum rules relating the coefficients of the high-frequency expansion of the self energy and the product of the self energy and Green function to thermodynamic expectation values. Comparison of non-crossing/one-crossing results to numerically exact quantum Monte-Carlo calculations shows that the consistency with sum rules provides a useful estimate of the reliability of the approximations. The sum rule violations are more pronounced, and therefore the quality of the non-crossing/one-crossing approximation is poorer, in situations with multiple orbitals and away from particle-hole symmetry but becomes less severe as the correlation strength increases. The one crossing approximation is markedly superior to the non-crossing approximation.
\end{abstract}
\date{\today}
\maketitle
\section{Introduction}
Electronic structure calculations of complex materials pose a highly non-trivial task: the interplay between charge, spin, orbital and lattice degrees of freedom can lead to striking many-body correlations challenging common band-structure approaches.\cite{Dagotto:2005} An important step towards a successful theoretical description of materials with strong electronic correlations was taken with the development of
dynamical mean-field theory (DMFT).\cite{Georges:1996,Kotliar:2006,Metzner:1989} DMFT provides a theoretical framework to account for 
correlations resulting from strong local interactions between electrons, by mapping the full problem onto a quantum impurity model with a self-consistently determined bath.

However, the solution of the quantum impurity model required in the DMFT method remains challenging.
The state-of-the-art methods to solve quantum impurity models make use of a stochastic sampling of diagrams in an imaginary-time expansion of the partition function.\cite{Rubtsov:2005,Werner:2006,Gull:2011} These 
``continuous time quantum Monte Carlo" (CT-QMC) approaches are numerically exact
and are 
widely used as ``impurity solvers" for DMFT calculations. However, 
these methods are computationally intensive (so that surveys of wide ranges of parameter space are often prohibitively expensive), are formulated on the Matsubara axis (so analytical continuation is required for real-frequency spectral information), and in many physically relevant cases suffer from a severe ``fermion sign problem''. This sign problem is severe if large clusters or many orbitals are simulated and may occur even in the single-site DMFT approximation,  
for multiorbital situations in which the local Green function 
is ``non-diagonal", i.e.~does not have a frequency independent eigenbasis. The latter situation
occurs generically in situations of low point symmetry, for example in the case of Co on Cu.\cite{Gorelov:2009}

In these situations, it is desirable to have a more economical way to solve the quantum impurity model by using approximations which give reasonably accurate results while keeping the computational cost at a minimum.
Self-consistent resummations of diagrams in the hybridization expansion are popular approximations because
they are based on the solution of integral equations rather than quantum field theories and 
the computational cost scales polynomially in the system size.\cite{Bickers:1987} Among various schemes, the non-crossing (NCA) and the one-crossing approximations (OCA) are frequently used in DMFT calculations.\cite{Pruschke:1993,Maier:2000,Zolfl:2000,Imai:2001,Okamoto:2008,Haule:2010,Eckstein:2010} 
These approaches 
may be 
formulated 
on the imaginary \cite{Haule:2010,Gull:2010} 
or on  the real
frequency axis using either Feynman's perturbation theory in a slave-boson representation with subsequent projection\cite{Coleman:1984,Haule:2001} or a perturbation theory based on contour integrals of ionic resolvents.\cite{Bickers:1987,Pruschke:1989}. Keldysh-contour formulations also have been studied.\cite{Eckstein:2010,Eckstein:2010b,Gull:2011b,Werner:2012} 
The ability to formulate the problem directly on the real axis or the Keldysh contour has the additional and very considerable advantage that analytical continuation is not necessary. 

The NCA/OCA and related approximations are obtained from summations of complete families of dressed skeleton diagrams and are therefore conserving approximations ($\Phi$-derivable).\cite{Grewe:1983,Bickers:1987,Pruschke:1993} This guarantees the equivalence of alternative representations of the partition function obtained by integrating thermodynamic derivatives -- a property which assures that {\it thermodynamic relations} are conserved within a given approximation. On the other hand, it is known that $\Phi$-derivability does {\it not} guarantee that sum rules (which connect frequency sums of dynamical quantities to equal-time correlation functions) and Fermi liquid relations (which connect thermodynamic derivatives or zero-frequency correlation functions and thermodynamic derivatives) are satisfied.\cite{Bickers:1987}
These have to be tested on a case-by-case basis. 

In the present article, we investigate the degree to which 
self-consistently resummed approximations such as 
the NCA and OCA respect 
sum rules 
relating the 
high-frequency expansion of the impurity Green's function $G(i\omega_n)$ [and self-energy $\Sigma(i\omega_n)$] 
to
thermodynamic expectation values of commutators of operators with the Hamiltonian
and to sum rules relating the Matsubara axis sum of the product of the Green function and self energy to the expectation value of the potential energy.
A comparison of NCA/OCA calculations to the results of numerically exact quantum Monte Carlo calculations shows that the degree of sum rule violation offers a straightforward and robust estimate of the quality of the approximation in the systems we tested. We also observe that the sum rule violation means that it is not possible to use sum rule techniques to estimate the high frequency tails needed for Fourier transformation.\cite{Knecht:2003,Comanac:2007}  
Although this paper presents explicit results only for  the NCA and OCA we observe that more involved conserving approximations\cite{Pruschke:1989,Haule:2001,Grewe:2008} can also be discussed within the framework developed here.

The quality of the NCA and related approximations was investigated previously but with a focus on the Fermi liquid properties.\cite{Muller-Hartmann:1984,Kuramoto:1984,Kroha:1997,Kirchner:2002,Kroha:2005} It was found that although the NCA gives qualitatively correct results for temperatures higher than the Kondo temperature, it develops a spurious non-analyticity at the Fermi energy at low temperatures.\cite{Muller-Hartmann:1984,Kuramoto:1984} To recover the correct Fermi liquid behavior, a considerably larger class of diagrams must be considered, as is the case in the conserving $T$-matrix approximation.\cite{Kroha:1997,Kirchner:2002} However, the application to dynamical mean field theory changes the focus from the details of the low frequency Fermi liquid behavior to the quality of the approximation at generic frequencies. There is thus a need for simple, robust estimators of the approximation quality. One of our aims in this paper is to show that sum rule violations provide such an estimator. 

The results of this paper are based on the imaginary-time formulation of the NCA/OCA for quantum impurity models with multiple orbitals. In Sec.~\ref{sec:formalism} we first review the formalism before we present in Sec.~\ref{sec:sum_rules} the above mentioned sum rules
and discuss the degree to which they are respected in the NCA and OCA.  In Sec.~\ref{sec:benchmark}, we directly benchmark the NCA/OCA against CT-QMC. From these tests, we conclude in Sec.~\ref{sec:conclusions} that the performance of the NCA/OCA is less satisfactory in situations with multiple orbitals and away from particle-hole symmetry. However, as expected, these approximations become better in situations where the hybridization is small compared to the interaction energy. Moreover, for the instances we have studied, the OCA provides a substantial improvement over the NCA. 

\section{Formalism}
\label{sec:formalism}
\subsection{Overview}
This section presents the formalism for the self-consistent resummation of the hybridization expansion in imaginary time, generalizing the scheme given in Ref.~\onlinecite{Gull:2010} to the multi-orbital case. 
We use a matrix notation which
makes the formalism independent of the details of the impurity model.  

\subsection{General impurity model}
We study impurity models of the  form
\begin{equation}
H=H_{\rm imp}+H_{\rm bath}+H_{\rm hyb},
\label{eq:H}
\end{equation}
where $H_{\rm imp}$ describes the (interacting) impurity electrons, $H_{\rm bath}$ the non-interacting bath electrons and $H_{\rm hyb}$ specifies the hybridization between the impurity and bath degrees of freedom. The mixing term has the general form
\begin{equation}
H_{\rm hyb}=\sum_{p,a}\left(V^a_{p}c_{p}^{\dag}d_a+{\rm h.c.}\right).
\end{equation}
Here, $d_a^{(\dag)}$ denotes the annihilation (creation) operators for the impurity electrons in spin-orbital $a$. $c_p^{(\dag)}$ describe the bath degrees of freedom which follow
\begin{equation}
H_{\rm bath}=\sum_p\varepsilon_pc^{\dag}_pc_p.
\end{equation}
In general, $p$ is a combined index including both the momentum and the internal quantum numbers such as spin. 

We expect that  $H_{\rm imp}$ is of the general form
\begin{equation}
H_{\rm imp}=\sum_{ab}E_{ab}d^\dagger_ad_b +H_{\rm int}.
\label{Himpdef}
\end{equation}
At this stage in the discussion we will not need to specify the interaction part $H_{\rm int}$.

We shall be interested in the imaginary-time d-electron Green function
\begin{equation}
G_{ab}(\tau)=-\left<{\mathcal T}_{\tau} d_a(\tau)d_b^{\dag}(0)\right>
\label{Gabdef}
\end{equation}
whose Fourier transform may be expressed in terms of the hybridization function matrix $\Delta$ and self energy matrix $\Sigma$ as
\begin{equation}
G(i\omega_n)=\left[i\omega_n\mathds{1}-E-\Delta(i\omega_n)-\Sigma(i\omega_n)\right]^{-1}.
\label{Gab2}
\end{equation}
The matrix $E$ specifies the single-particle levels of the impurity and is given in Eq.~\eqref{Himpdef}.

The hybridization function
\begin{equation}
\Delta_{ab}(\tau)=-\frac{1}{N_{\rm bath}}\sum_p{V_p^a}^*V_p^b\langle{\mathcal T}_{\tau} c_p(\tau)c_p^{\dag}(0)\rangle.
\label{hybdef}
\end{equation}
arises from integrating out the bath electrons. In terms of the bath dispersion $\varepsilon_p$, it has the explicit representation\cite{Gull:2011}
\begin{equation*}
\Delta_{ab}(\tau)=\frac{1}{N_{\rm bath}}\sum_p\frac{{V_p^a}^*V_p^b}{e^{\varepsilon_p\beta}+1}\times
\begin{cases}
-e^{-\varepsilon_p(\tau-\beta)}, & 0<\tau<\beta;\\
e^{-\varepsilon_p\tau}, & -\beta<\tau<0.
\end{cases}
\end{equation*} 
which has the property  that its Fourier transform $\Delta(\omega)$ vanishes as $|\omega|\rightarrow\infty$.

The self energy expresses the effect of the interaction terms $H_{\rm int}$ on the d-electron dynamics. 

\subsection{Hybridization expansion}
The starting point for the approximations discussed in this paper is the hybridization expansion of the partition function.\cite{Werner:2006,Werner:2006b,Gull:2011} It uses the interaction representation with respect to the hybridization $H_{\rm hyb}$: 
\begin{eqnarray}
Z&=&{\rm Tr}\left[e^{-\beta H_0}{\mathcal T}e^{-\int_0^{\beta}d\tau V({\tau})}\right]\nonumber\\
&=&\sum_{k=0}^{\infty}\int_0^{\beta}\!\tau_1\dots\int_{\tau_{k-1}}^{\beta}\!d\tau_k{\rm Tr}[e^{-\beta H_0}e^{\tau_k H_0}(-V)\nonumber\\
&&\dots e^{-(\tau_2-\tau_1)H_0}(-V)e^{-\tau_1H_0}],
\label{eq:Z}
\end{eqnarray}
where $V=H_{\rm hyb}$ and $H_0=H_{\rm imp}+H_{\rm bath}$.
Only even powers of this expansion with equal number of creation and annihilation operators contribute. After separating the bath and impurity operators one can integrate out the contribution from the bath degrees of freedom. Using Wick's theorem 
the contributions of the bath electrons 
can be written in terms of the hybridization function defined in Eq.~\ref{hybdef}.

Collecting terms of the same order in $\Delta_{ab}(\tau)$, the expansion of the partition function takes the final form\cite{Werner:2006,Gull:2010} 
\begin{eqnarray}
Z=Z_{\rm bath}\sum_{k}\iiint d\tau_1\dots d\tau_k'\sum_{j_1\dots j_k}\sum_{j_1'\dots j_k'}{\rm Tr}[{\mathcal T}_{\tau}e^{-\beta H_{\rm imp}}\nonumber\\
\times d_{j_k}(\tau_k)d_{j_k'}^{\dag}(\tau_k')\dots d_{j_1}(\tau_1)d_{j_1'}^{\dag}(\tau_1')]{\rm det}\ {\bs \Delta},
\label{eq:Zexp}
\end{eqnarray}
where ${\bs \Delta}$ is a $k\times k$ matrix with entries $\Delta_{lm}=\Delta_{j_lj_m}(\tau_l-\tau_m')$.
It is possible to use a 
Monte Carlo algorithm to 
evaluate the series stochastically, thereby computing observables like the Green's function numerically exactly.\cite{Werner:2006,Werner:2006b,Gull:2011} 
It is also possible to provide an approximate evaluation by resumming particular subsets of  terms in a self-consistent manner. Two well-known examples are the non-crossing (NCA)  and one-crossing (OCA) approximations. 
\subsection{The non-crossing approximation (NCA)}
\begin{figure}
\includegraphics[width=0.8\linewidth]{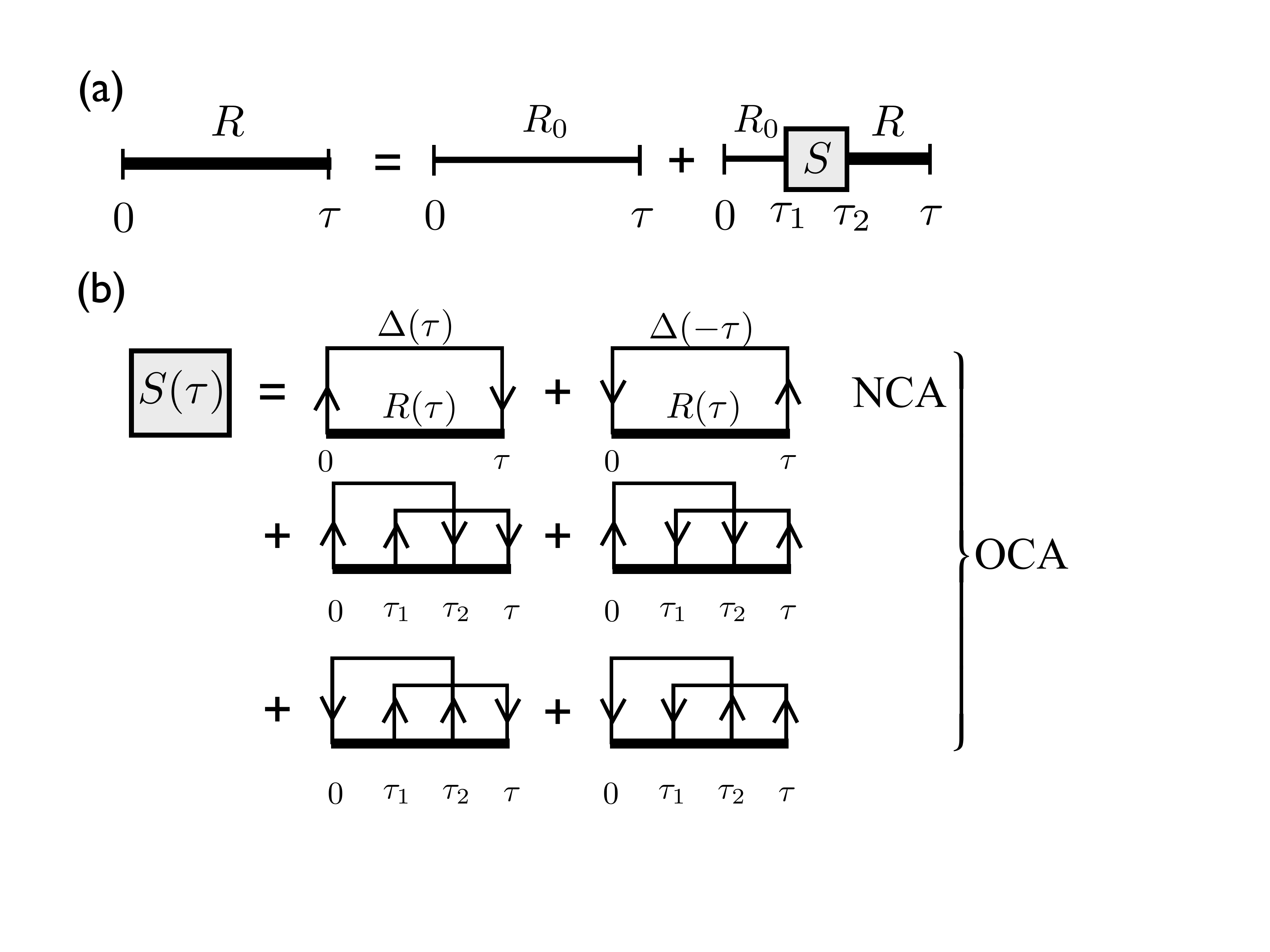}
\caption{(a) The Dyson equation for the self-consistent local propagator $R(\tau)$. (b) The self-energy $S(\tau)$ of the local propagator in the NCA and OCA.}
\label{fig:diagrams}
\end{figure}

The non-crossing approximation (NCA) is a resummation of all the terms in Eq.~\eqref{eq:Zexp} which have non-crossing hybridization lines. It can be obtained by considering the $k=0,1$ terms in Eq.~\eqref{eq:Zexp} but with a dressed propagator of the local eigenstates, $R(\tau)$. $R(\tau)$ is a $N\times N$-matrix, where $N$ is the dimension of the local Hilbert space. It fulfills the following Dyson equation in imaginary time, see Fig.~\ref{fig:diagrams}(a):
\begin{eqnarray}
R(\tau)=R_0(\tau)\!+\!\int_0^{\tau}\!\!d\tau_2\!\!\int_0^{\tau_2}\!\!d\tau_1R(\tau-\tau_2)S(\tau_2-\tau_1)R_0(\tau_1)%\nonumber\\
\label{eq:Dyson}
\end{eqnarray}
Here, the bare propagator $R_0(\tau)$ is given by
\begin{equation}
R_0(\tau)=e^{-\tau H_{\rm imp}}.
\end{equation}
By construction, $R(\tau)$ is only defined for $0<\tau<\beta$. The $N\times N$-matrix $S(\tau)$ corresponds to the ``self-energy" of the local propagator $R(\tau)$. In the NCA, it is given by $S(\tau)=S^0(\tau)$ where
\begin{equation}
S^0(\tau)=\sum_{ab}\left[d_aR(\tau)d_b^{\dag}\Delta_{ba}(-\tau)-d_a^{\dag}R(\tau)d_b\Delta_{ab}(\tau)\right].
\label{eq:SNCA}
\end{equation}
The creation and annihilation operators $d_a^{(\dag)}$ in Eq.~\eqref{eq:SNCA} should be interpreted as their corresponding matrix representations in the local Hilbert space. Equation~\eqref{eq:SNCA} has the diagrammatic representation shown in the first line of Fig.~\ref{fig:diagrams}(b). It has to be solved self-consistently together with Eq.~\eqref{eq:Dyson}, in practice this is typically  done by iteration.

Once a self-consistent solution is found, physical quantities are calculated from $R(\tau)$. For example, the partition function is given by
\begin{equation}
Z={\rm Tr}\left[R(\beta)\right].
\end{equation}
Furthermore, static (thermodynamic) expectation values are readily computed,
\begin{equation}
\left\langle O\right\rangle_{\rm NCA}=\frac{1}{Z}{\rm Tr}\left[R(\beta)O\right],
\label{eq:TDexp}
\end{equation}
where $O$ is an arbitrary local operator. 

Dynamical (imaginary time) quantities are calculated in a similar way. The most important dynamical quantity for the following discussion is the physical single-particle Green's function of the impurity site:
\begin{equation}
G_{ab}(\tau_2-\tau_1)=-\langle{\mathcal T}_{\tau}d_a(\tau_2)d_b^{\dag}(\tau_1)\rangle.
\end{equation}
Within NCA, it is obtained as
\begin{equation}
G_{ab}(\tau)=
\begin{cases}
-{\rm Tr}\left[R(\beta-\tau)d_aR(\tau)d_b^{\dag}\right]/Z,& 0<\tau<\beta;\\
{\rm Tr}\left[R(\beta+\tau)d_b^{\dag}R(-\tau)d_a\right]/Z,&-\beta<\tau<0.
\end{cases}
\label{eq:GNCA}
\end{equation}

The non-crossing approximation is a conserving approximation.\cite{Bickers:1987} In particular, there exists a Luttinger-Ward functional $\Phi[R,\Delta]$ from which the local eigenstate self-energy $S(\tau)$ as well as the impurity Green's function $G(\tau)$ are obtained by a functional derivative:
\begin{eqnarray}
S_{nm}(\tau)&=&\frac{\delta\Phi[R,\Delta]}{\delta R_{mn}(\beta-\tau)},\label{eq:S_Phi}\\
G_{ab}(\tau)&=&\frac{1}{Z}\frac{\delta\Phi[R,\Delta]}{\delta\Delta_{ba}(\beta-\tau)}\label{eq:G_Phi}.
\end{eqnarray}
In the NCA, the Luttinger-Ward functional is $\Phi[R,\Delta]=\Phi^0[R,\Delta]$ where\cite{Haule:2001}
\begin{equation}
\Phi^0[R,\Delta]\!=\!-\sum_{a,b}\int_0^{\beta}\!d\tau {\rm Tr}\left[R(\beta-\tau)d_a R(\tau) d_b^{\dag}\right]\Delta_{ba}(\beta-\tau).
\end{equation}
Using $\Phi^0[R,\Delta]$ in Eqs.~\eqref{eq:S_Phi} and \eqref{eq:G_Phi} one recovers Eq.~\eqref{eq:SNCA} for the self-energy and Eq.~\eqref{eq:GNCA} for the impurity Green's function in the NCA. 

\subsection{The one-crossing approximation (OCA)}
The NCA was originally developed for the infinite-$U$ single-orbital Anderson model where the approximation works well.\cite{Bickers:1987} For finite $U$, the NCA shows severe problems because it neglects exchange contributions.\cite{Pruschke:1989,Haule:2001} There are many different schemes which improve on the NCA by including diagrams with crossing hybridization lines.\cite{Grewe:2008} We will discuss the simplest such generalization and will refer to it as the one-crossing approximation (OCA).\cite{Haule:2010,Eckstein:2010} Other names used in the literature for the same approximation include ``enhanced NCA" (ENCA),\cite{Schmitt:2009,Grewe:2008,Grewe:2009} ``finite $U$ NCA" (UNCA),\cite{Haule:2001} or ``first-order one-crossing approximation" (1st-order OCA).\cite{Gull:2010}

The OCA is obtained from the expansion Eq.~\eqref{eq:Zexp} by additionally including the $k=2$ terms with crossing hybridization lines. Again, the approximation is made self-consistent by taking a dressed propagator $R(\tau)$ which fulfills the Dyson equation Eq.~\eqref{eq:Dyson}, in analogy to the NCA. However, as illustrated in Fig.~\ref{fig:diagrams}(b), the OCA includes additional exchange diagrams for the self-energy, $S(\tau)=S^0(\tau)+S^1(\tau)$.
\begin{widetext}
The
OCA is also a conserving approximation. The
Luttinger-Ward functional 
is obtained from the NCA functional by adding
an additional contribution, $\Phi[R,\Delta]=\Phi^0[R,\Delta]+\Phi^1[R,\Delta]$, where\cite{Kotliar:2006}
\begin{equation}
\Phi^1[R,\Delta]=-\sum_{\alpha\beta\gamma\delta}\int_{0}^{\beta}\!\!d\tau_3\int_{0}^{\tau_3}\!\!d\tau_2\int_0^{\tau_2}\!\!d\tau_1{\rm Tr}\left[R(\beta-\tau_3)d_{\delta}^{\dag}R(\tau_3-\tau_2)d_{\gamma}R(\tau_2-\tau_1)d_{\beta}R(\tau_1)d_{\alpha}^{\dag}\right]\Delta_{\alpha\gamma}(\beta-\tau_2)\Delta_{\delta\beta}(\tau_3-\tau_1).
\end{equation}
$S^1(\tau)$ can now be obtained from Eq.~\eqref{eq:S_Phi} by replacing $\Phi[R,\Delta]$ by $\Phi^1[R,\Delta]$.
The explicit expression for $S^1(\tau)$ is given by
\begin{eqnarray}
S^{1}(\tau)&=&-\sum_{abcd}\int_0^{\tau}d\tau_2\int_0^{\tau_2}d\tau_1\Big[d_d^{\dag}R(\tau-\tau_2)d_c^{\dag}R(\tau_2-\tau_1)d_bR(\tau_1)d_a\Delta_{ca}(\tau_2)\Delta_{db}(\tau-\tau_1)\nonumber\\
&&+d_dR(\tau-\tau_2)d_c^{\dag}R(\tau_2-\tau_1)d_b^{\dag}R(\tau_1)d_a\Delta_{ca}(\tau_2)\Delta_{bd}(\beta-\tau+\tau_1)\nonumber\\
&&+d_d^{\dag}R(\tau-\tau_2)d_cR(\tau_2-\tau_1)d_bR(\tau_1)d_a^{\dag}\Delta_{ac}(\beta-\tau_2)\Delta_{db}(\tau-\tau_1)\nonumber\\
&&+d_dR(\tau-\tau_2)d_cR(\tau_2-\tau_1)d_b^{\dag}R(\tau_1)d_a^{\dag}\Delta_{ac}(\beta-\tau_2)\Delta_{bd}(\beta-\tau+\tau_1)\Big].
\label{eq:SOCA}
\end{eqnarray}
Similarly, the impurity Green's function $G(\tau)$ acquires an additional contribution to Eq.~\eqref{eq:GNCA}. The full expression is 
\begin{eqnarray}
G_{ab}(\tau)&=&-{\rm Tr}[R(\beta-\tau)d_aR(\tau)d_b^{\dag}]/Z\nonumber\\
&&-\sum_{cd}\int_0^{\tau}d\tau_1\int_{\tau}^{\beta}d\tau_2{\rm Tr}\left[R(\beta-\tau_2)d_cR(\tau_2-\tau)d_aR(\tau-\tau_1)d_d^{\dag}R(\tau_1)d_b^{\dag}\Delta_{dc}(\beta-\tau_2+\tau_1)\right]/Z\nonumber\\
&&-\sum_{cd}\int_0^{\tau}d\tau_1\int_{\tau}^{\beta}d\tau_2{\rm Tr}\left[R(\beta-\tau_2)d_c^{\dag}R(\tau_2-\tau)d_aR(\tau-\tau_1)d_dR(\tau_1)d_b^{\dag}\Delta_{cd}(\tau_2-\tau_1)\right]/Z,
\label{eq:GOCA}
\end{eqnarray}
which can be obtained from $\Phi[R,\Delta]$ by the functional derivative Eq.~\eqref{eq:G_Phi}.
\end{widetext}
\section{Sum rules}
\label{sec:sum_rules}
\subsection{Overview}
In the following, we study the 
degree to which 
the NCA/OCA 
respects sum rules. 
In Sec.~\ref{sec:high_freq} we test the degree to which the NCA and OCA respect the  relations, known from the exact theory,  between  the coefficients of the high-frequency expansion of $G(i\omega_n)$ and independently known thermodynamic expectation values. In Sec.~\ref{sec:Epot} we investigate the sum rule for the potential energy. 
We present general arguments showing that neither the NCA nor the OCA fulfills the sum rules. 
In Sec.~\ref{sec:num}, we present numerical results for these sum rule violations.%%

\subsection{High-frequency expansion }
\label{sec:high_freq}
The high-frequency expansion of the impurity Green's function in Matsubara frequency space is 
given by (here we omit the matrix indices of $G$ etc for ease of writing)
\begin{equation}
G(i\omega_n)=\int_0^{\beta}d\tau G(\tau)e^{i\omega_n\tau}=\sum_{k\geq1}\frac{c_k}{(i\omega_n)^k}.
\label{eq:Gomega}
\end{equation}
Note that $(c_1)_{ab}=\delta_{ab}$ which insures that the single-particle spectral function is normalized to 1.
The self energy and hybridization function have similar high-frequency expansions
\begin{equation}
\Sigma(i\omega_n)=\sum_{k\geq 0}\frac{\Sigma_k}{(i\omega_n)^k},
\label{eq:sigmahigh}
\end{equation}
and
\begin{equation}
\Delta(i\omega_n)=\sum_{k\geq 1}\frac{\Delta_k}{(i\omega_n)^k}.
\label{eq:deltahigh}
\end{equation}
Note that the moments $\Delta_k$ are known {\it a priori} and that the hybridization function is defined so that $\Delta_{k=0}=0$.  $\Sigma_{k=0}$ gives the Hartree shift of the levels of the impurity model specified by the matrix $E$. Use of the Kramers-Kronig relation implied by the causality of the self energy implies
\begin{equation}
\Sigma_{k=1}=\int\frac{d\omega}{\pi} {\rm Im}\, \Sigma^{\rm ret}(\omega)
\label{eq:sigmasumrule}
\end{equation}
so that $\Sigma_{k=1}$ contains information about the interaction-induced dynamics. 

Comparison of Eqs.~\eqref{Gab2} and ~\eqref{eq:Gomega} shows that
\begin{eqnarray}
c_2&=&E+\Sigma_{0}
\label{c2relation} \\
c_3&=&\left(E+\Sigma_{0}\right)^2+\Delta_1+\Sigma_1.
\label{c3relation}
\end{eqnarray}

A relation between the  moments $c_k$ in the high-frequency expansion of $G$ and the
discontinuities in 
the derivatives of the Green function at $\tau=0$ follows from repeated integration by parts of Eq.~\ref{Gabdef}:
\begin{equation}
c_k=(-1)^{k}\left[G^{(k-1)}(0^+)-G^{(k-1)}(0^-)\right]
\label{ckdef}
\end{equation}
(here $G^{(k)}$ denotes the $k^{th}$ derivative of $G$). The time derivatives may also be obtained by expanding the Heisenberg equation of motion $\mathcal{O}(\tau)=e^{-H\tau}\mathcal{O}e^{H\tau}$ for small times  \cite{Knecht:2003,Comanac:2007} 
\begin{equation}
G_{ab}^{(k)}(0^+)-G_{ab}^{(k)}(0^-)=-\left\langle\left\{[\underbrace{H,[H,\dots[H}_{k {\rm times}},d_{a}]\dots]],d_b^{\dag}\right\}\right\rangle.
\label{eq:jumps}
\end{equation}
Detailed expressions for the commutators for general models  are available in the literature (see, e.g.~Ref.~\onlinecite{Potthoff:1997,Wang11}).

Equation~\eqref{eq:jumps} via Eqs.~\eqref{ckdef} and ~\eqref{eq:Gomega} provides an exact relation between the moments in the high frequency expansion to the equal-time expectation value of $k$-fold commutators of the exact Hamiltonian with fermion operators.
In an approximate solution of the impurity problem, the left and right side of Eq.~\eqref{eq:jumps} are in general not equal. Therefore, comparing both sides order by order provides a test for the quality of an approximation. Within the NCA/OCA, we find that the relation \eqref{eq:jumps} is in general violated for $k>1$. Below we explicitly discuss the first three terms in the high-frequency expansion. 
\subsubsection{$k=1$:}
To determine the zeroth order we compute the discontinuity of the Green's function at $\tau=0$. Within NCA/OCA, one obtains
\begin{eqnarray*}
G_{ab}(0^+)&=&-{\rm Tr}\left[R(\beta)d_ad_b^{\dag}\right]/Z,\\
G_{ab}(0^-)&=&{\rm Tr}\left[R(\beta)d_b^{\dag}d_a\right]/Z
\end{eqnarray*}
which results in
\begin{equation}
\left(c_1^{\rm NCA/OCA}\right)_{ab}=\langle\{d_a,d_b^{\dag}\}\rangle_{\rm NCA/OCA}=\delta_{ab}.
\label{eq:c1}
\end{equation}
Here, $\langle\dots\rangle_{\rm NCA/OCA}$ denotes the thermodynamic expectation value within the NCA or OCA, as given by Eq.~\eqref{eq:TDexp}. Equation~\eqref{eq:c1} is consistent with the relations Eq.~\eqref{eq:jumps} and guarantees, e.g.,~that the single-particle spectral function is correctly normalized.
\subsubsection{$k=2$:}
For the next-higher term we first consider the NCA. Evaluating the first derivative of Eq.~\eqref{eq:GNCA} at $\tau=0^{\pm}$ we obtain
\begin{eqnarray*}
\left(c_2^{\rm NCA}\right)_{ab}=\frac{1}{Z}\left\{{\rm Tr}\left[R'(\beta)d_ad_b^{\dag}\right]-{\rm Tr}\left[R'(\beta)d_b^{\dag}d_a\right]\right.\nonumber\\
+\left.{\rm Tr}\left[R(\beta)d_b^{\dag}R'(0)d_a\right]-{\rm Tr}\left[R(\beta)d_aR'(0)d_b^{\dag}\right]\right\}.
\end{eqnarray*}
The above expression can be further simplified by using $R'(0)=-H_{\rm imp}$ and Eq.~\eqref{eq:Dyson} to obtain $R'(\beta)$. After some algebra, one finds the following form
\begin{equation}
\left(c_2^{\rm NCA}\right)_{ab}=-\left\langle\left\{\left[H,d_a\right],d_b^{\dag}\right\}\right\rangle_{\rm NCA}+\epsilon^{0}_{ab}.
\label{eq:c2NCA}
\end{equation}
where for consistency the commutator should be evaluated within the NCA approximation as indicated by the subscript NCA. 

In deriving Eq.~\ref{eq:c2NCA} we used the fact that for the impurity model Eq.~\eqref{eq:H}, the $k=1$ anti-commutator is independent of the hybridization and the bath degrees of freedom:
\begin{equation}
\left\{\left[H_{\rm imp},d_a\right],d_b^{\dag}\right\}=\left\{\left[H,d_a\right],d_b^{\dag}\right\}.
\end{equation}
This allowed us to replace $H_{\rm imp}$ by the full Hamiltonian $H$ in Eq.~\eqref{eq:c2NCA}.
The second term appearing in Eq.~\eqref{eq:c2NCA} is 
the $l=0$ member of a family of expressions given by the general formula
\begin{eqnarray}
\epsilon^l_{ab}=\frac{1}{Z}\int_0^{\beta}\!\!d\tau\left\{{\rm Tr}\left[R(\beta-\tau)S^l(\tau)d_ad_b^{\dag}\right]\right.\nonumber\\
-\left.{\rm Tr}\left[S^l(\tau)R(\beta-\tau)d_b^{\dag}d_a\right]\right\}
\label{eq:E}
\end{eqnarray}
for $S^{0,1}(\tau)$ given by Eq.~\eqref{eq:SNCA} and Eq.~\eqref{eq:SOCA}.
Comparison of  Eq.~\eqref{eq:c2NCA} to Eq.~\eqref{eq:jumps} makes it clear that a non-zero $\epsilon_{ab}^0$ indicates that in the NCA the high-frequency tail of the Green function is not given by the general commutator expression evaluated within the same theory.  
Because $\epsilon_{ab}^0$ involves an integral over $S^0(\tau)$, it is proportional to $V^2$ for small $V$ [see Eq.~\eqref{eq:SNCA}]. We find that except for the particle-hole symmetric limit, $\epsilon_{ab}^0$ is indeed non-zero, which reflects the fact that the impurity Green function in the NCA is exact only in zeroth order in the hybridization strength.

A similar evaluation for the OCA Green's function [Eq.~\eqref{eq:GOCA}] yields an analogous result
\begin{equation}
\left(c_2^{\rm OCA}\right)_{ab}=-\left\langle\left\{\left[H,d_a\right],d_b^{\dag}\right\}\right\rangle_{\rm OCA}+\epsilon^{1}_{ab}.
\label{eq:c2OCA}
\end{equation}
with the error $\epsilon_{ab}^1$  now given by Eq.~\eqref{eq:E} with $l=1$. $\epsilon_{ab}^1$ involves an integral over the exchange contributions $S^1(\tau)$ of the self-energy which is proportional to $V^4$ for small $V$ [see Eq.~\eqref{eq:SOCA}] instead of $V^2$ in the NCA. The inconsistency in the OCA is therefore considerably smaller than in the NCA at small $V$ (i.e. large $U$).

\subsubsection{$k=3$:}
The $k=3$ term in the high-frequency expansion of the impurity Green function is given by
\begin{equation}
(c_3)_{ab}=\left\langle\left\{\left[H,\left[H,d_a\right]\right],d_b^{\dag}\right\}\right\rangle.
\end{equation}
Similar algebra as for the $k=2$ term, but too lengthy to reproduce here, shows that the $k=3$ and all the higher frequency moments of the Green function suffer from similar sum rule violations as the $k=2$ term.

The failure of the NCA/OCA to reproduce the relation between the high frequency tails of the Green function and the commutators means that the relations between the high frequency components of the self energy and expectations values of commutators are similarly in error. This is in particular true for the first two moments, the Hartree shift $\Sigma_0$ and the $1/\omega_n$ term $\Sigma_1$ which can be obtained from Eqs.~\eqref{c2relation} and \eqref{c3relation}. By comparing the NCA/OCA to numerically exact CT-QMC data, we will later argue (see Secs.~\ref{sec:numS1} and \ref{sec:benchmark} as well as Figs.~\ref{fig:Sigma1_sl},\ref{fig:Sigma1_sf},\ref{fig:Sigma_sl} and \ref{fig:Sigma_sf}) that the sum rule violation for $\Sigma_1$ is a good diagnostic for the quality of the approximation at general frequencies.

\subsection{Potential energy sum rule}
\label{sec:Epot}
A related 
error appears  in the sum rule for the potential energy. On the one hand, the NCA/OCA allows to directly compute the static expectation value
\begin{equation}
\tilde{E}_{\rm pot}^{\rm stat}=\langle H_{\rm int}\rangle
\label{eq:Epot(1)tilde}
\end{equation}
where $H_{\rm int}=H_{\rm imp}-\sum_{ab}E_{ab}d_a^{\dag}d_b$ denotes the local interaction Hamiltonian. On the other hand, the potential energy can also be obtained from a sum rule if the impurity Green's function and the self-energy are known by evaluating the following expression:\cite{Fetter:2003,Georges:1996} 
\begin{equation}
\tilde{E}_{\rm pot}^{\rm sum}=\frac{1}{2\beta}\sum_n{\rm Tr}\left[\Sigma(i\omega_n)G(i\omega_n)\right]e^{i\omega_n 0^+}.
\label{eq:Epot(2)tilde}
\end{equation}
Here, $\rm Tr$ denotes the trace over the spin and orbital degrees of freedom. We find that within NCA/OCA, $\tilde{E}_{\rm pot}^{\rm stat}\neq \tilde{E}_{\rm pot}^{\rm sum}$ in general. As we show in the following, the difference between the two expressions has a similar origin as the inconsistency in the high-frequency expansion. Indeed, the basis for Eq.~\eqref{eq:Epot(1)tilde} is the relation 
\begin{eqnarray}
{\rm Re}\left\{{\rm Tr}\left[G'(0^-)\right]\right\}&=&-\frac{1}{2}\sum_a\left(\langle[H,d_a^{\dag}]d_a\rangle+\langle d_a^{\dag}[d_a,H]\rangle\right)\nonumber\\
&=&-\langle H_{\rm hyb}\rangle -{\rm Tr}[E \rho]-2\langle H_{\rm int}\rangle.
\label{eq:Gp}
\end{eqnarray}
In the last line we introduced $\rho_{ab}=\langle d_a^{\dag}d_b\rangle$. Equation~\eqref{eq:Gp} is an exact relation which can be derived from the definition of the imaginary-time Green's function (or the equation of motion), similar to Eq.~\eqref{eq:jumps}. Writing the left-hand side in frequency space, using the Dyson Eq.~\eqref{Gab2} for the impurity Green's function and the relations
\begin{eqnarray}
\rho_{ab}=\langle d_a^{\dag}d_b \rangle&=&\frac{1}{\beta}\sum_nG_{ab}(i\omega_n)e^{i\omega_n0^+},\\
E_{\rm hyb}=\langle H_{\rm hyb}\rangle&=&{\rm Tr}\frac{1}{\beta}\sum_n\Delta(i\omega_n)G(i\omega_n)e^{i\omega_n 0^+},
\label{eq:E_hyb}
\end{eqnarray}
one can solve Eq.~\eqref{eq:Gp} for $\langle H_{\rm int}\rangle$ which yields the expression Eq.~\eqref{eq:Epot(1)tilde}. As we have shown in the previous section, the first derivative of $G(\tau)$ at $\tau=0^\pm$ is not simply obtained from thermodynamic expectation values within NCA/OCA. Hence, the relation Eq.~\eqref{eq:Gp} is in general (except for vanishing hybridization) not satisfied within these approximations and therefore also $\tilde{E}_{\rm pot}^{\rm stat}\neq \tilde{E}_{\rm pot}^{\rm sum}$.

Because the NCA and OCA preserve  particle-hole symmetry, it is convenient to bring the expressions Eqs.~\eqref{eq:Epot(1)tilde} and \eqref{eq:Epot(2)tilde} into a form which respects this symmetry (if present). For simplicity, we assume $E_{ab}=-\mu\delta_{ab}$ and define\begin{eqnarray}
E_{\rm pot}^{\rm stat}&=&\tilde{E}_{\rm pot}^{\rm stat}-\mu_0 n,\label{eq:Epot(1)}\\
E_{\rm pot}^{\rm sum}&=&\tilde{E}_{\rm pot}^{\rm sum}-\frac{1}{4}{\rm Tr}\,\Sigma_0-\frac{\mu_0 n}{2}.\label{eq:Epot(2)}
\end{eqnarray}
Here, $\mu_0$ is the value of $\mu$ for which the impurity is half filled. If the impurity model is particle-hole symmetric, the above expressions respect this symmetry as well.
It is then natural to quantify the sum rule violation by the ratio
\begin{equation}
\left|\frac{\Delta E_{\rm pot}}{\mu_0}\right|=\left|\frac{\tilde{E}_{\rm pot}^{\rm stat}-\tilde{E}_{\rm pot}^{\rm sum}+\frac{1}{4}{\rm Tr}\, \epsilon^{l}}{\mu_0}\right|.
\label{eq:DeltaE}
\end{equation}
where $\Delta E_{\rm pot}=E_{\rm pot}^{\rm stat}-E_{\rm pot}^{\rm sum}$ and $\epsilon^l$ with $l=0$ (NCA) or $l=1$ (OCA) is given in Eq.~\eqref{eq:E}. If the NCA/OCA works well, one expects $\left|\Delta E_{\rm pot}/\mu_0\right|\ll 1$ which we indeed observed by direct comparison with numerically exact CT-QMC data, see Sec.~\ref{sec:benchmark}. From the examples studied, we found that one can use Eq.~\eqref{eq:DeltaE} as a tool to estimate the quality of the approximation.

\subsection{Numerical results}
\label{sec:num}
\subsubsection{Two level quantum dot model}
\label{sec:QD}
The numerical results presented in the following sections are obtained for
a model for a two level quantum dot (impurity with two orbitals) with asymmetric coupling to two leads (bath degrees of freedom). The model has been studied in Ref.~\onlinecite{WangMillis:2010} in view of potential quantum critical points related to the occupancy switching of the two levels. Here, we use it to illustrate the internal inconsistencies one can encounter in the NCA/OCA and to benchmark our NCA/OCA calculations against the CT-QMC results of Ref.~\onlinecite{WangMillis:2010}. 

The two orbitals are labeled with the index $\alpha={\rm n, w}$, distinguishing between narrow (n) and wide (w) level. We study both spinless and spinful impurity electrons interacting via an interorbital repulsion $U$ and coupled to the bath via orbital dependent parameters $V_{\alpha}$ 
(note that in neither case is an intra-orbital interaction included). 
The spinless version of the model takes the form
\begin{equation}
H_{\rm sl}=Un_{\rm n}n_{\rm w}-\mu\sum_{\alpha}n_{\alpha}+\sum_{p,\alpha}V_{\alpha}\left(c_p^{\dag}d_{\alpha}+d_{\alpha}^{\dag}c_p\right)+H_{\rm bath}
\label{eq:Hsl}
\end{equation}
with $n_{\alpha}=d_{\alpha}^{\dag}d_{\alpha}$ and $\mu$ the chemical potential. The spinful version is the same but with spin indices added:
\begin{equation}
H_{\rm sf}=Un_{\rm n}n_{\rm w}-\mu\sum_{\alpha}n_{\alpha}+\sum_{p,\alpha,\sigma}V_{\alpha}\left(c_{p\sigma}^{\dag}d_{\alpha\sigma}+d_{\alpha\sigma}^{\dag}c_{p\sigma}\right)+H_{\rm bath}
\label{eq:Hsf}
\end{equation}
where $\sigma=\uparrow$, $\downarrow$ labels the spin and $n_{\alpha}$ now $=\sum_{\sigma}d_{\alpha\sigma}^{\dag}d_{\alpha\sigma}$.

We assume that the bath degrees of freedom are described by a broad and featureless band with a semi-circular density of states of width $W=4t$:
\begin{equation}
\rho(\varepsilon)=\frac{\sqrt{4t^2-\varepsilon^2}}{2t^2\pi}.
\end{equation}
The coupling $V_{\alpha}$ of the two quantum dot levels to the leads introduces a broadening of the levels. In the non-interacting limit 
for dots with energy levels close to the center of the band and weak hybridization (as compared to $W$), the broadening of the two levels is given by 
\begin{equation}
\Gamma_{\rm n}=\pi|V_{\rm n}|^2\rho(0)\quad {\rm and}\quad \Gamma_{\rm w}=\pi|V_{\rm w}|^2\rho(0).
\end{equation}
Throughout this article we assume $\Gamma_{\rm n}=0.04 t$ and $\Gamma_{\rm w}=0.25t$ which is much smaller than the band width $W=4t$ of the bath electrons. The width of the broader level is chosen as the unit of energy, i.e.~$\Gamma_{\rm w}=1$. In these units, the level broadening of the narrow level is $\Gamma_{\rm n}=0.16$ and the band width is $W=16$.

Performing the commutators shows that for these models the coefficient $c_2^{\alpha}$ controlling the $1/\omega_n^2$ decay of the Green function for orbital $\alpha$ is ($\bar{\alpha}$ denotes the other orbital)
\begin{equation}
c_2^{\alpha}=U\left<\hat{n}_{\bar{\alpha}}\right>-\mu=Un_{\bar{\alpha}}-\mu
\label{c2model}
\end{equation}
from which we obtain
\begin{equation}
\Sigma_0^{\alpha}=Un_{\bar{\alpha}}.
\label{eq:Sigma0}
\end{equation}
for the Hartree shift.
The coefficient $\Sigma_1^{\alpha}$ giving the $1/\omega_n$ term in the self energy is
\begin{equation}
\Sigma_1^{\alpha}=U^2\left(\left<\hat{n}_{\bar{\alpha}}^2\right>-\left(\left<\hat{n}_{\bar{\alpha}}\right>\right)^2\right).
\label{sigma1model}
\end{equation}
In the spinless model $\hat{n}_{\bar{\alpha}}^2=\hat{n}_{\bar{\alpha}}$ so the expression reduces to 
\begin{equation}
\Sigma_{1,sl}^{\alpha}=U^2\left(n_{\bar{\alpha}}-n_{\bar{\alpha}}^2\right)
\label{sigma1model}
\end{equation}
but in the spinful model the expectation value of $\hat{n}_{\alpha}^2$ enters.

\subsubsection{Numerical results for Hartree shift $\Sigma_0$}
\begin{figure}
\includegraphics[width=0.49\linewidth]{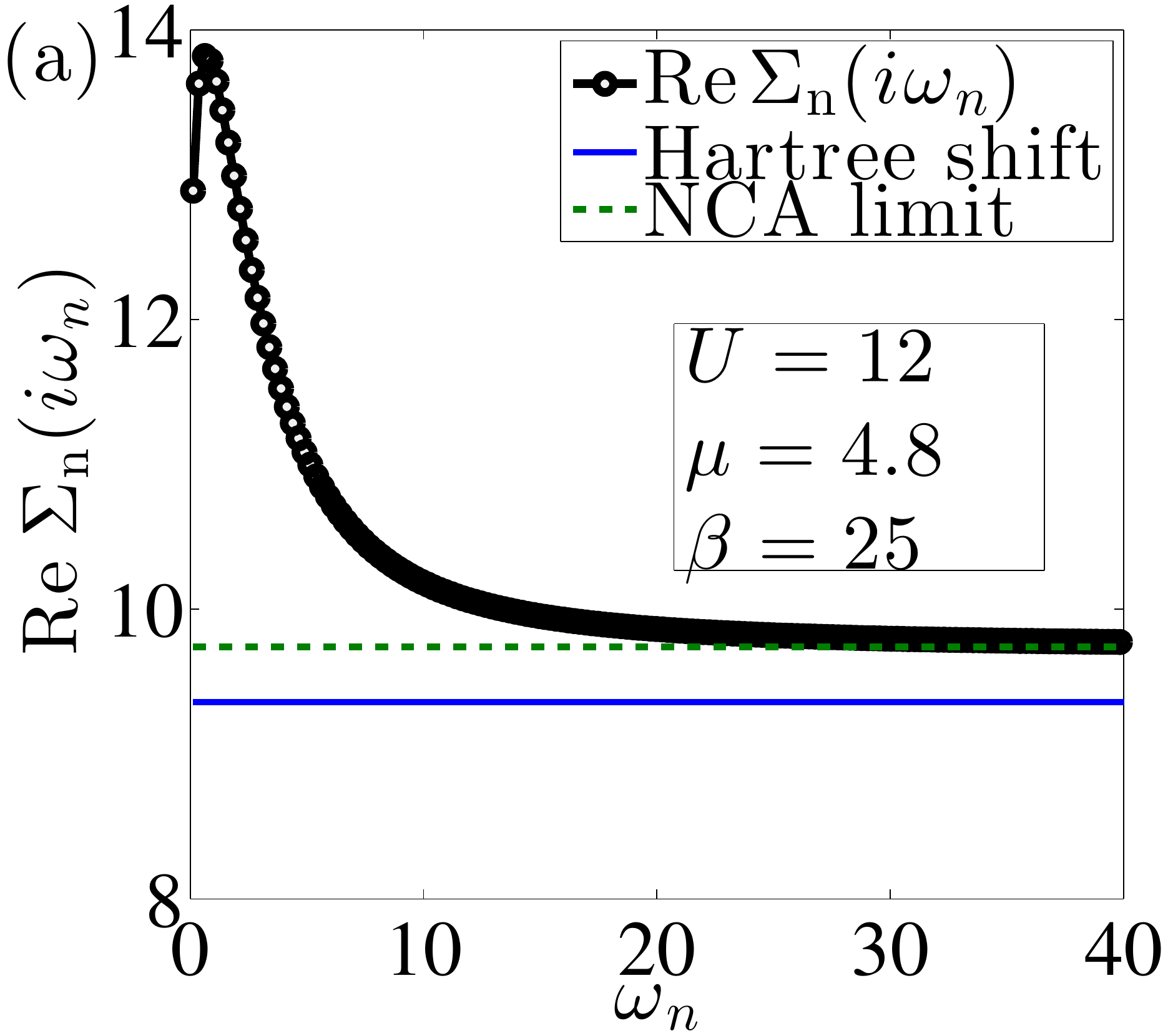}
\includegraphics[width=0.50\linewidth]{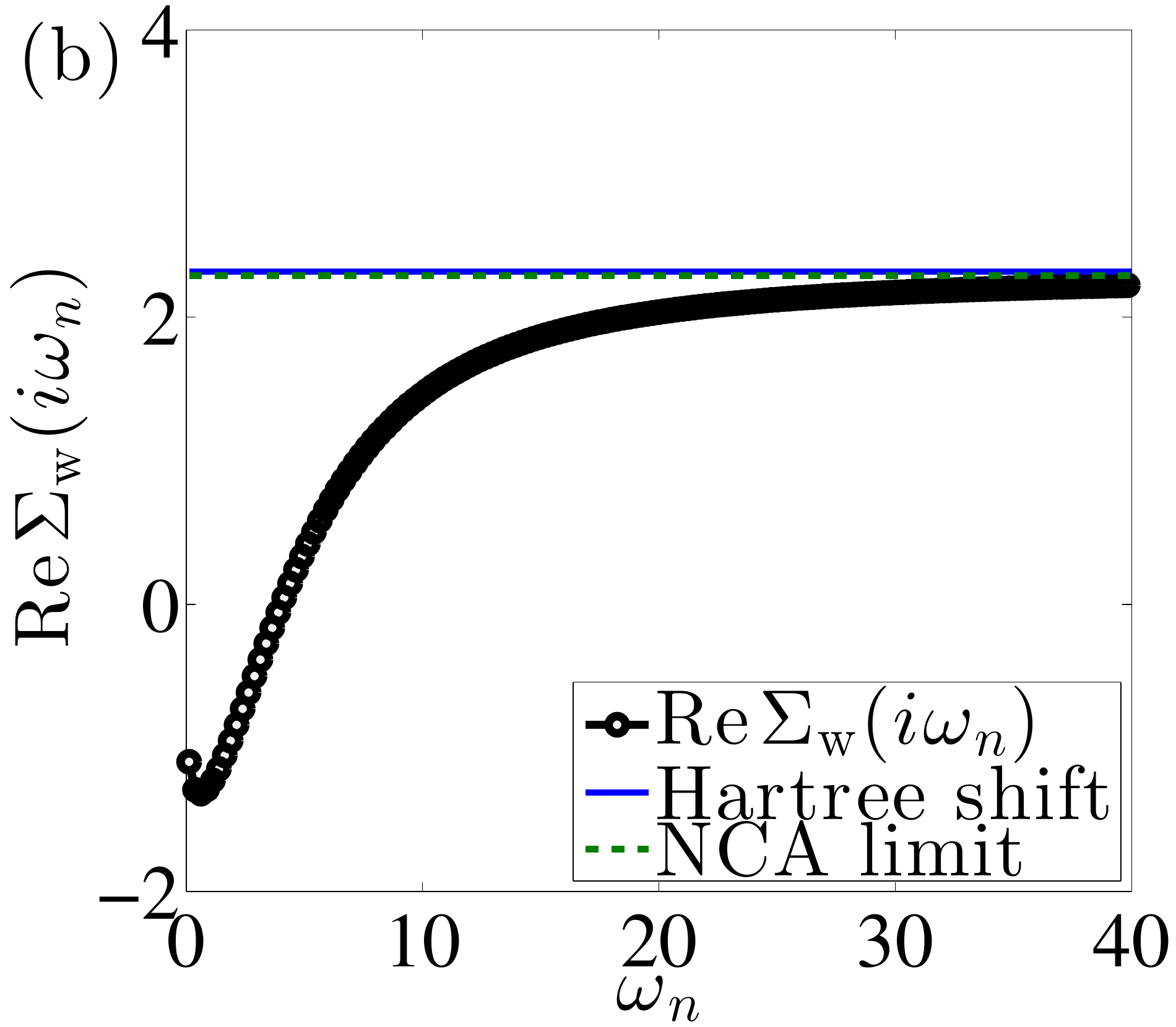}
\caption{(Color online) High-frequency behavior of the self-energy in the NCA for the spinless model for (a) the narrow level and (b) the wide level. The (blue) solid line represents the Hartree shift expected from the sum rule. The (green) dashed line represents the high-frequency limit obtained from the NCA equations with  $U=12$, $\mu=4.8$ and $\beta=25$.}
\label{fig:Sigma0_sl_NCA}
\end{figure}
\begin{figure}
\includegraphics[width=0.51\linewidth]{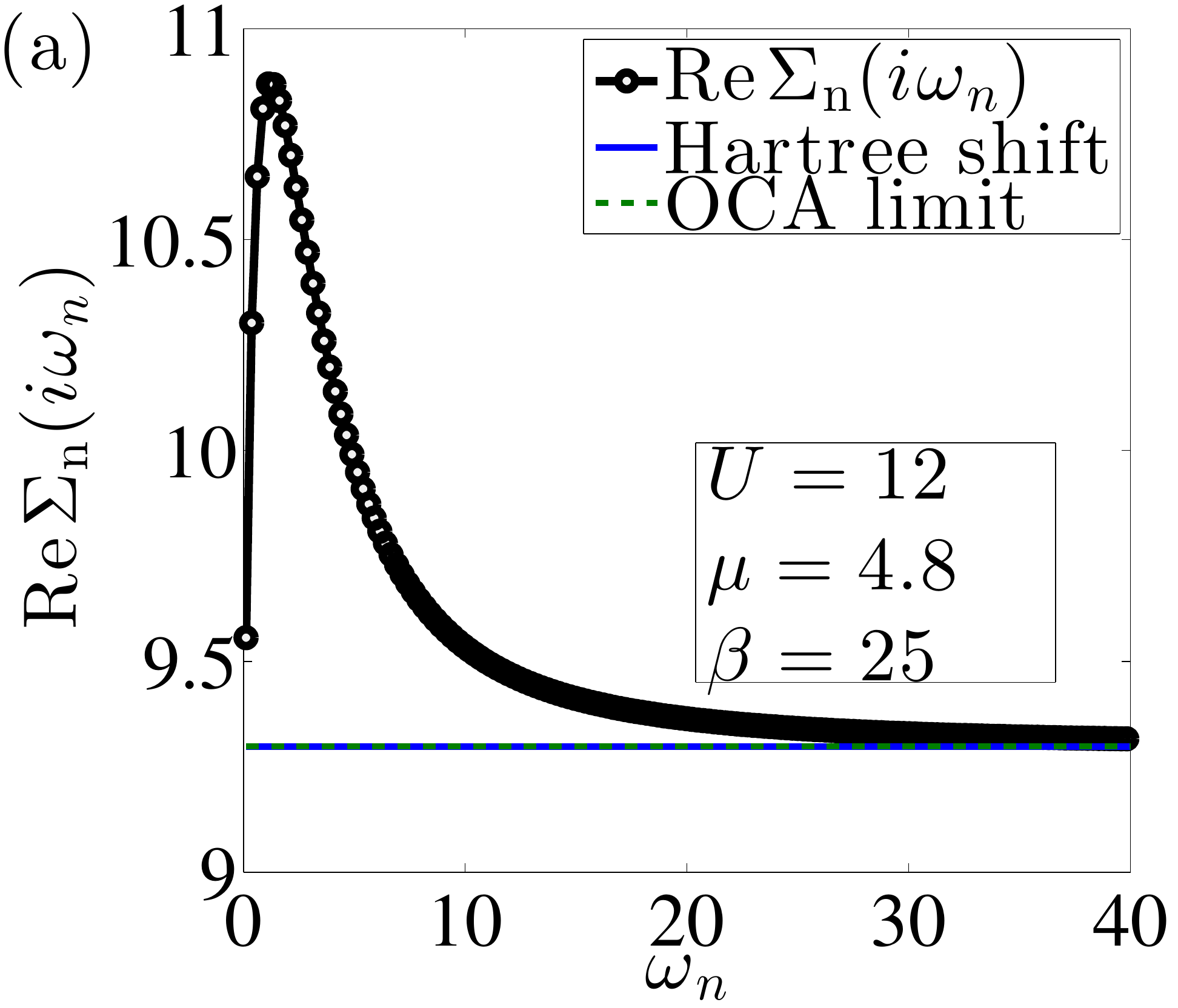}
\includegraphics[width=0.48\linewidth]{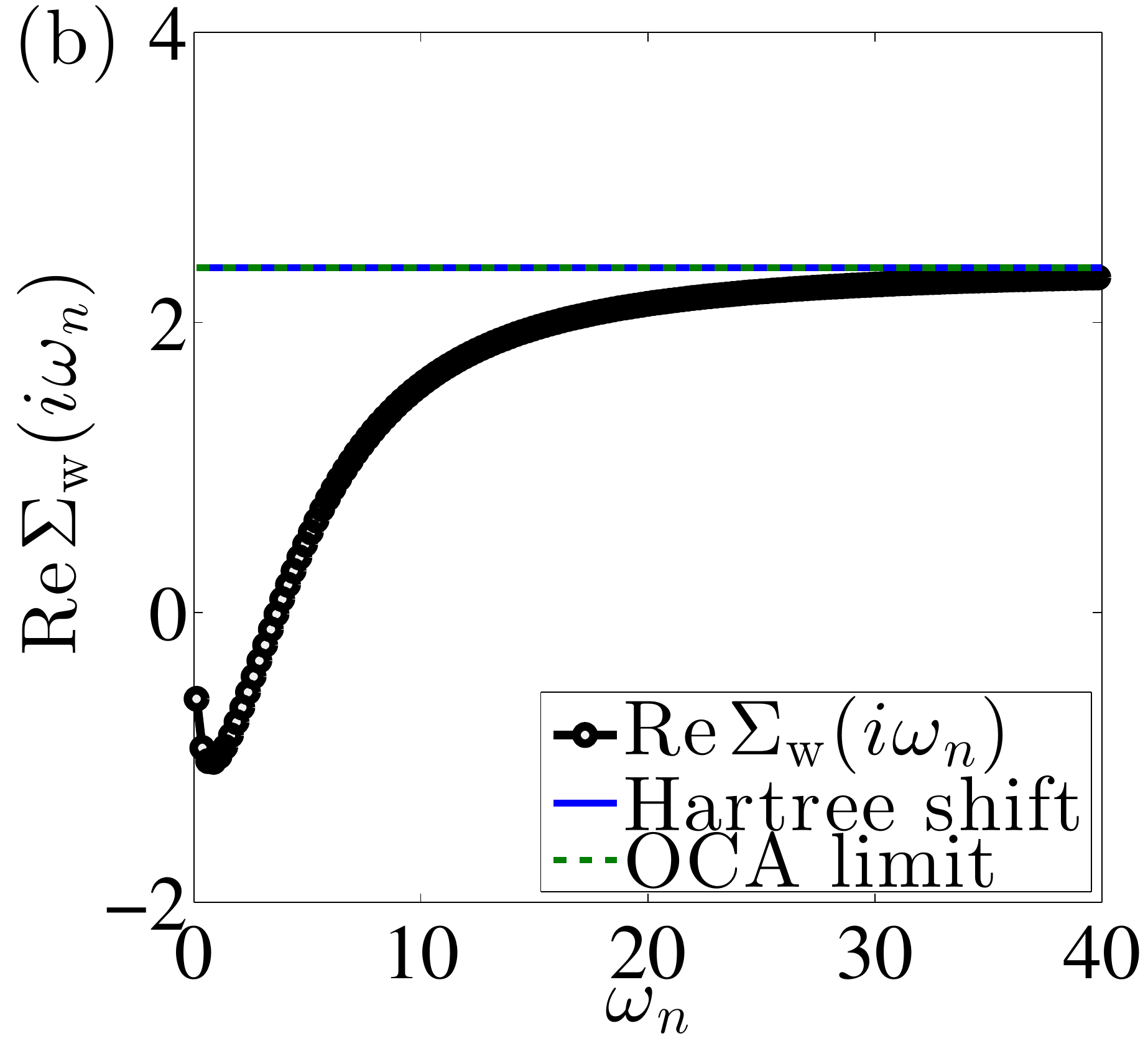}
\caption{(Color online) High-frequency behavior of the self-energy in the OCA for the spinless model for (a) the narrow level and (b) the wide level. The (blue) solid line represents the Hartree shift expected from the sum rule. The (green) dashed line represents the high-frequency limit obtained from the OCA equations which is indistinguishable from the Hartree shift within the resolution of the graph. The following parameters have been used: $U=12$, $\mu=4.8$ and $\beta=25$.}
\label{fig:Sigma0_sl_OCA}
\end{figure}
In the following we present numerical results for the Hartree shift $\Sigma_0$.
Particle-hole symmetry protects the value $\Sigma_0=\mu_0$ where $\mu_{0,sl}=U/2$ for the spinless and $\mu_{0,sf}=U$ for the spinful model. This symmetry protection is respected by the NCA and OCA so we focus on results away from the particle-hole symmetric limit. 

We first consider the spinless model Eq.~\eqref{eq:Hsl}.  
Figures~\ref{fig:Sigma0_sl_NCA} and \ref{fig:Sigma0_sl_OCA} show the real part of the self-energy for the narrow ($\alpha={\rm n}$) and the wide ($\alpha={\rm w}$) level as obtained in the NCA and OCA, respectively. For these calculations, we have used a large interorbital interaction $U=12$ and have fixed $\mu=4.8$ and $\beta=25$. For these parameters, the total filling $n=n_{\rm n}+n_{\rm w}$ is slightly below half-filling, $n=1$, and the broader level is preferably occupied, $n_{\rm w}>n_{\rm n}$. In the NCA (Fig.~\ref{fig:Sigma0_sl_NCA}), the difference between the value for Hartree shift expected from the sum rule, Eq.~\eqref{eq:Sigma0}, and the actual high-frequency limit of the NCA self-energy is noticeable. The discrepancy is clearly visible for the narrow level but quite small for the wide level. In the OCA (Fig.~\ref{fig:Sigma0_sl_OCA}), a distinction is not resolved within numerical precision.

\begin{figure}
\includegraphics[width=0.49\linewidth]{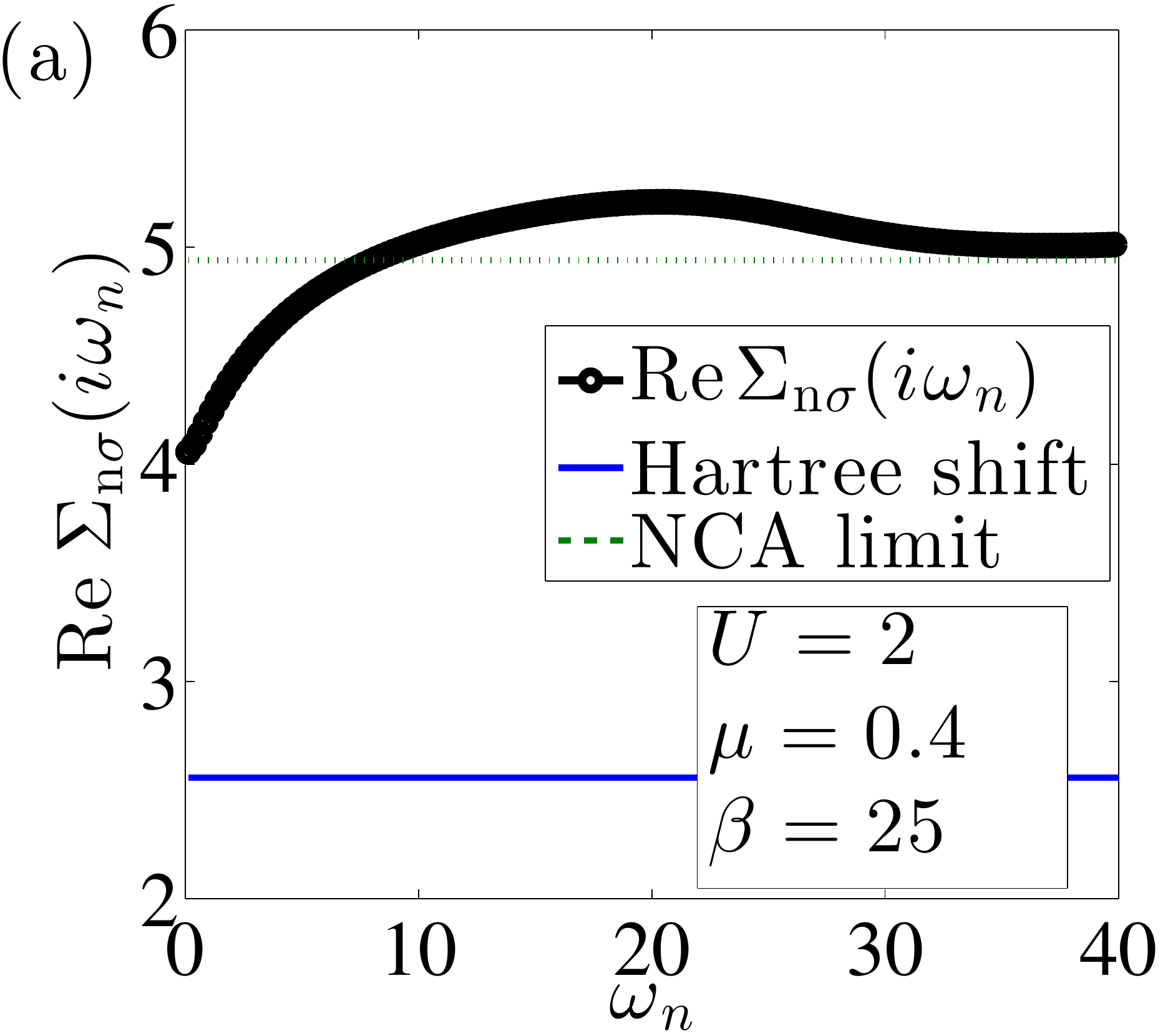}
\includegraphics[width=0.49\linewidth]{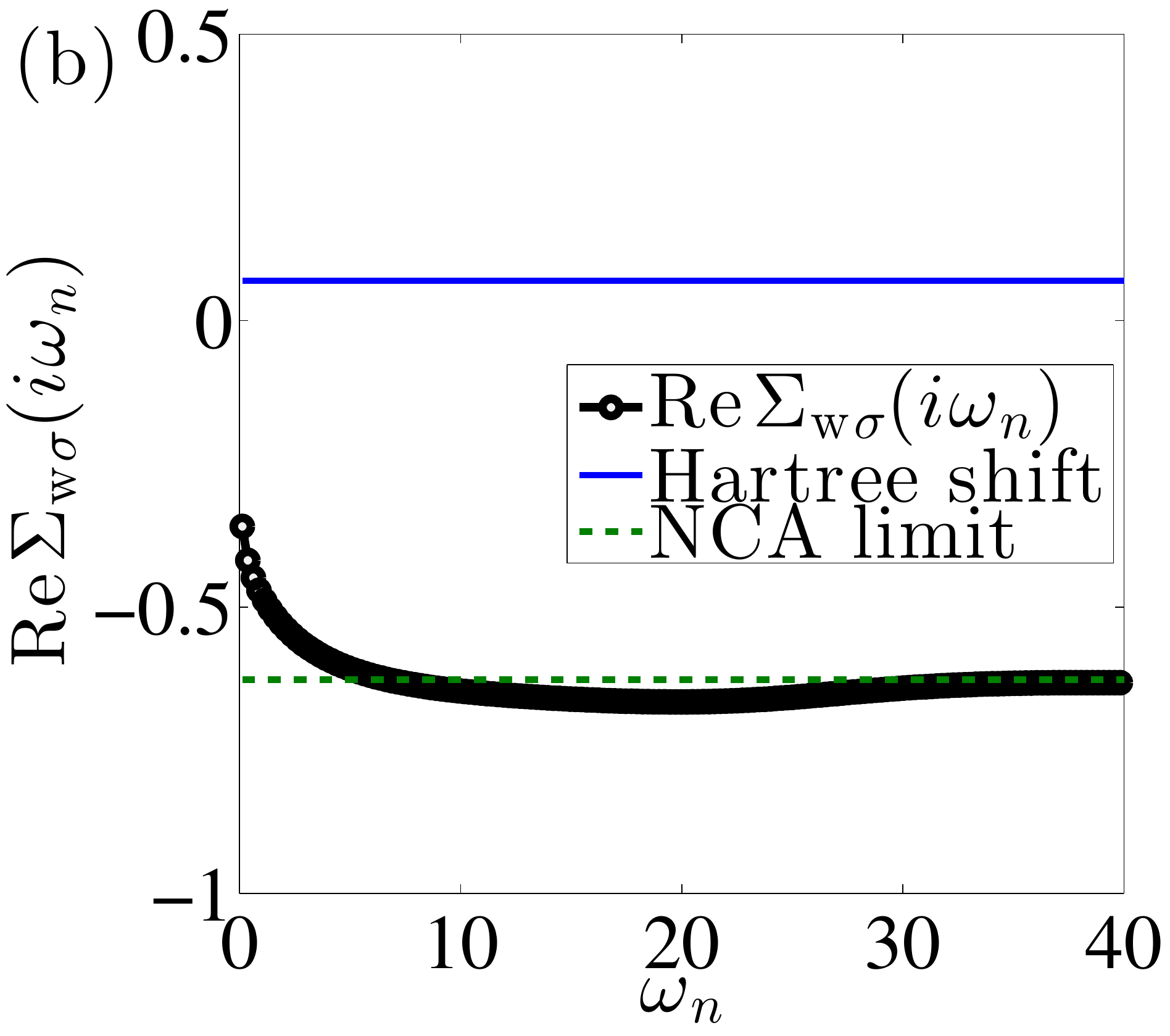}
\caption{(Color online) Real part of the self-energy for the spinful two-orbital model within the NCA for (a) the narrow and (b) the wide level with spin $\sigma$. The (blue) solid line represents the Hartree shift expected from the sum rule. The (green) dashed line represents the high-frequency limit obtained from the NCA equations.}
\label{fig:Sigma0_sf_NCA}
\end{figure}

\begin{figure}
\includegraphics[width=0.51\linewidth]{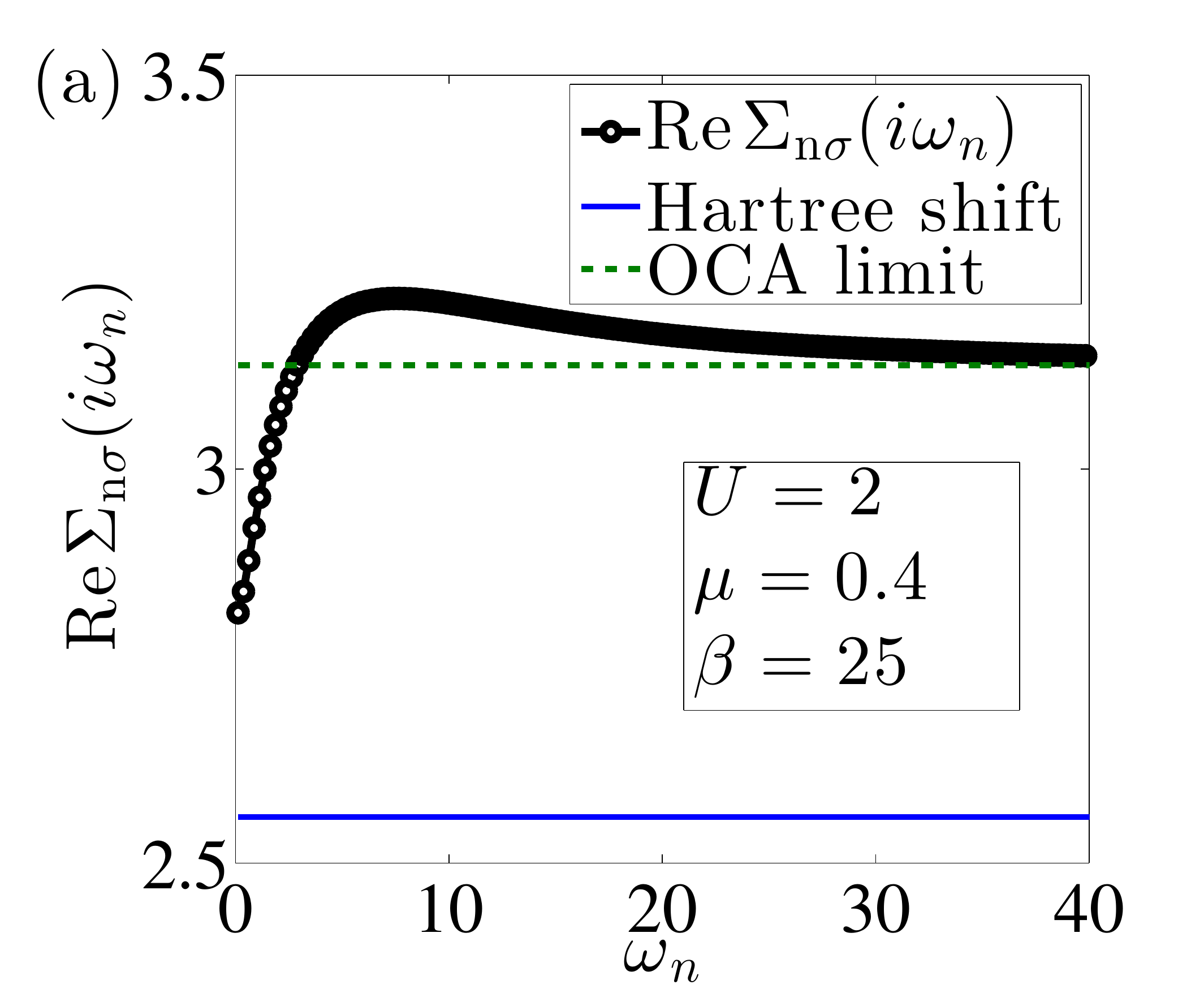}
\includegraphics[width=0.48\linewidth]{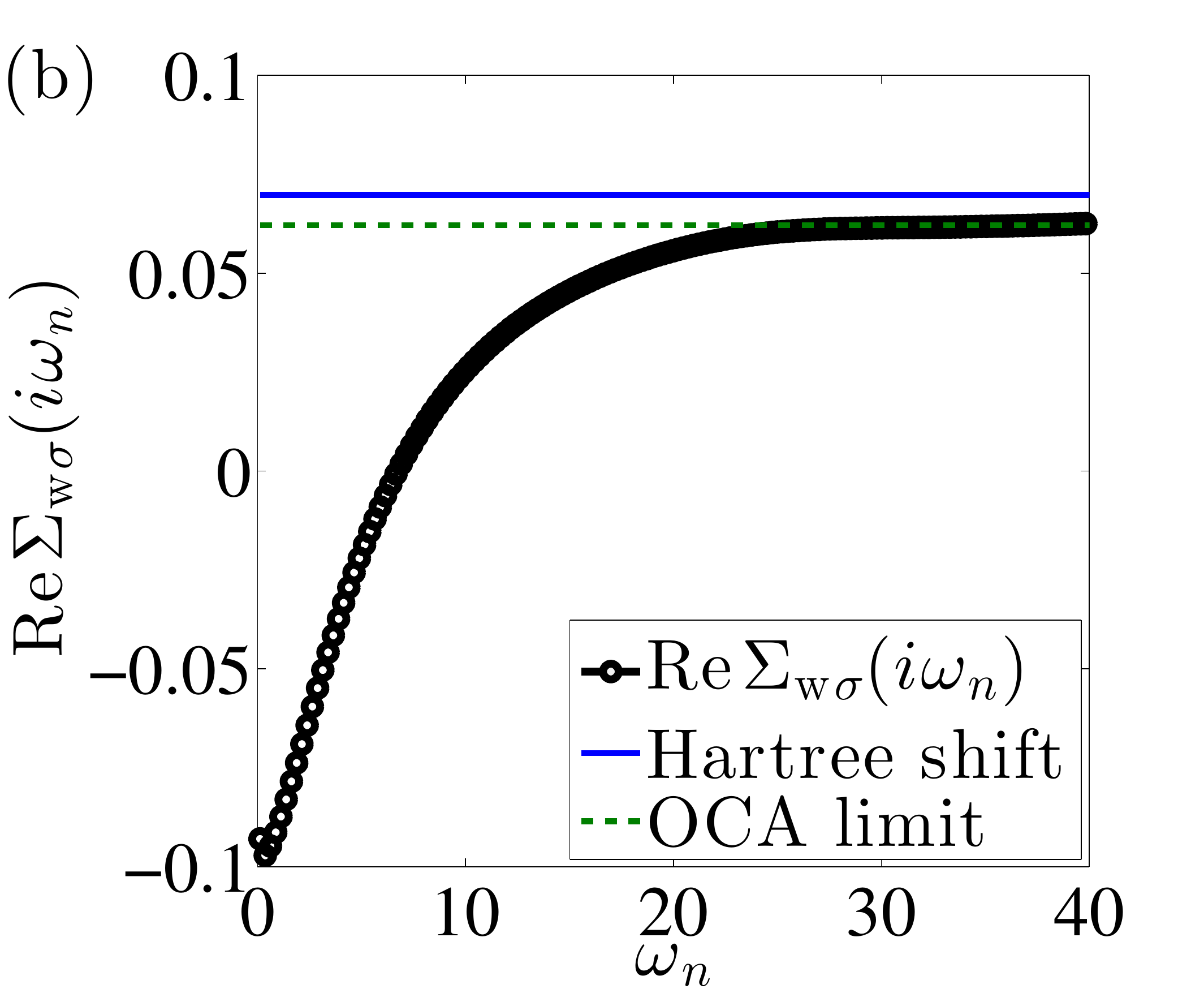}
\caption{(Color online) Real part of the self-energy for the spinful two-orbital model within the OCA for (a) the narrow and (b) the wide level with spin $\sigma$. The (blue) solid line represents the Hartree shift expected from the sum rule. The (green) dashed line represents the high-frequency limit obtained from the OCA equations.}
\label{fig:Sigma0_sf_OCA}
\end{figure}

We next consider the spinful model Eq.~\eqref{eq:Hsf}. 
Again, we compare the value from Eq.~\eqref{eq:Sigma0} to the high-frequency limit of the real part of the self-energy in the NCA/OCA, see Figs.~\ref{fig:Sigma0_sf_NCA} and \ref{fig:Sigma0_sf_OCA}. For a given orbital, the self-energy is identical for the two spin components and we show only one. The parameters were chosen as $U=2$, $\mu=0.4$ and $\beta=25$. We note that the discrepancy between the value of the Hartree shift from the sum rule and the actual high-frequency limit is now manifest in both approximations. Moreover, Fig.~\ref{fig:Sigma0_sf_NCA}(b) shows that also negative (unphysical) values for the high-frequency limit are possible in the NCA/OCA. 

\begin{figure}
\includegraphics[width=0.75\linewidth]{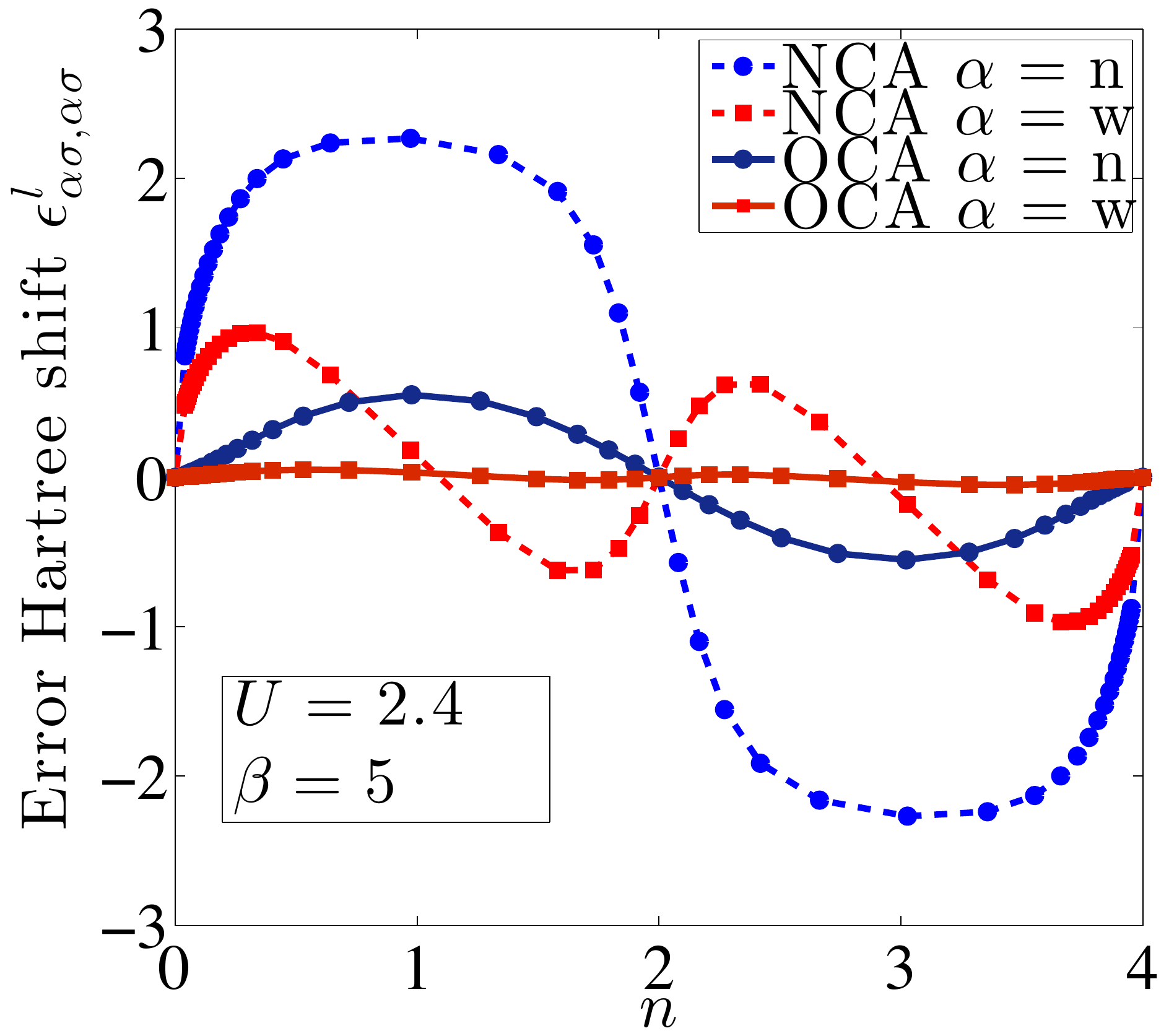}
\caption{(Color online) The term  $\epsilon_{\alpha\sigma,\alpha\sigma}^{l}$ [Eq.~\eqref{eq:E}] which quantifies the sum rule violation in the Hartree shift as function of the electron density $n$ for $U=2.4$ and $\beta=5$ as obtained from the NCA and OCA in the spinful model. Note the clear improvement of the OCA as compared to the NCA.}
\label{fig:E}
\end{figure}

In Fig.~\ref{fig:E} we finally show the dependence of the sum-rule violation term  $\epsilon_{\alpha\sigma,\alpha\sigma}^{l}$ [Eq.~\eqref{eq:E}] on the electronic density $n$ for the spinful model. As mentioned earlier, $\epsilon_{\alpha\sigma,\alpha\sigma}^{l}$ vanishes at particle-hole symmetry ($n=2$) and also approaches zero in the limits $n\rightarrow 0$ and $n\rightarrow 4$. Notice the clear improvement of the OCA over the NCA.

\subsubsection{Numerical results for $\Sigma_1$}
\label{sec:numS1}
The next higher moment in the high-frequency expansion of the self-energy is given by the coefficient $\Sigma_1$ determining the asymptotic $1/\omega_n$ behavior. The sum rule for $\Sigma_1$ is generally violated even at particle-hole symmetry and in the following we present results for this case.
 
\begin{figure}
\includegraphics[width=0.9\linewidth]{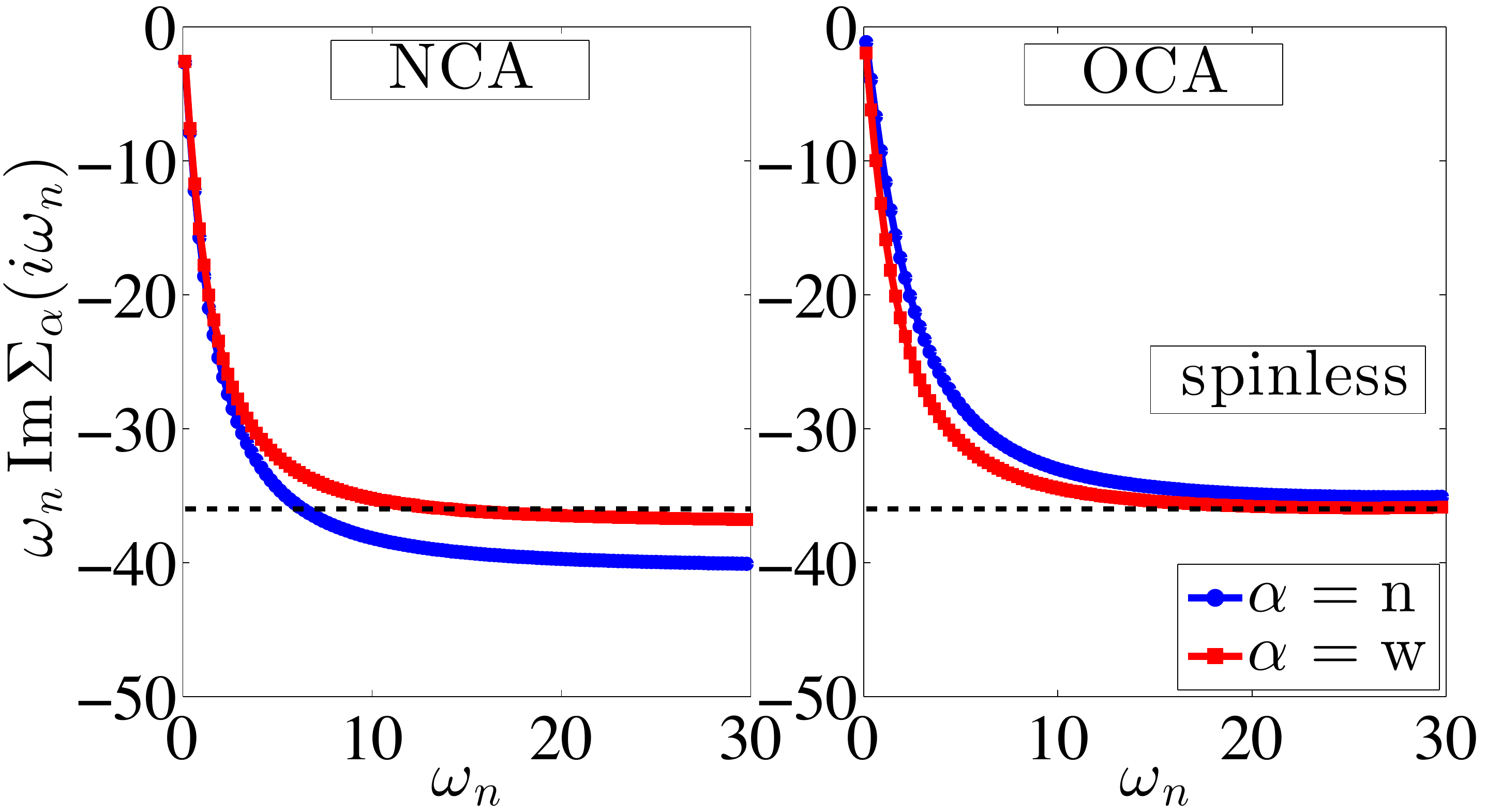}
\caption{(Color online) Comparison between $\omega_n {\rm Im}\, \Sigma_{\alpha}(i\omega_n)$ and the asymptotic value if the sum rule for the coefficient $\Sigma_1$ [Eq.~\eqref{sigma1model}] was fulfilled (dashed line). Results are obtained within NCA (left panel) and OCA (right panel) for the spinless model at particle-hole symmetry with $U=12$, $\mu=4.5$ and $\beta=25$. For these parameters, a direct comparison between NCA/OCA and CT-QMC is provided in Fig.~\ref{fig:Sigma_sl}.}
\label{fig:Sigma1_sl}
\end{figure}

Figure~\ref{fig:Sigma1_sl} compares $\omega_n{\rm Im}\,\Sigma_{\alpha}(i\omega_n)$ to the asymptotic value expected from Eq.~\eqref{sigma1model} for the spinless model with large interactions $U=12$. In the NCA, we find that the sum rule for the wide level is satisfied within a few percent while for the narrow level it is roughly 10-15\%. In the OCA, the sum rule is satisfied within the precision of the graph.

\begin{figure}
\includegraphics[width=0.9\linewidth]{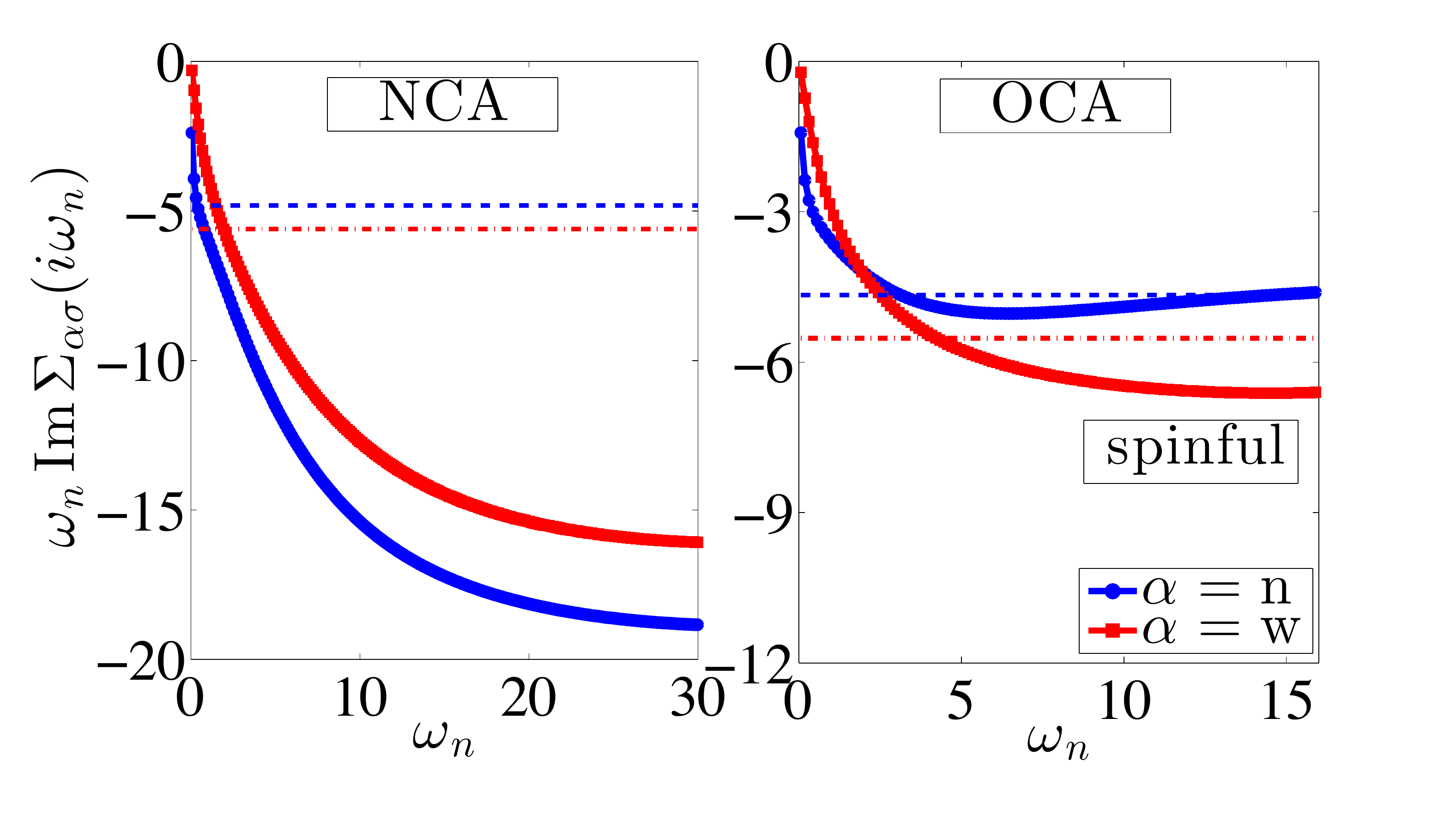}
\caption{(Color online) Comparison between $\omega_n {\rm Im}\, \Sigma_{\alpha\sigma}(i\omega_n)$ and the asymptotic value if the sum rule for the coefficient $\Sigma_1$ was fulfilled (dashed line for the narrow level $\alpha={\rm n}$ and dashed-dotted line for the wide level $\alpha={\rm w}$). Results are obtained within NCA (left panel) and OCA (right panel) for the spinful model at particle-hole symmetry with $U=2.4$, $\mu=2.4$ and $\beta=50$. For these parameters, a direct comparison between NCA/OCA and CT-QMC is provided in Fig.~\ref{fig:Sigma_sf}.}
\label{fig:Sigma1_sf}
\end{figure}

Figure~\ref{fig:Sigma1_sf} shows the same analysis for the spinful model at moderately strong interactions $U=2.4$. While the sum rule violation in the NCA is rather striking, the OCA clearly improves leading to an overall agreement of 10-15\%.
\subsubsection{Numerical results for the potential energy}
\begin{figure}
\includegraphics[width=0.8\linewidth]{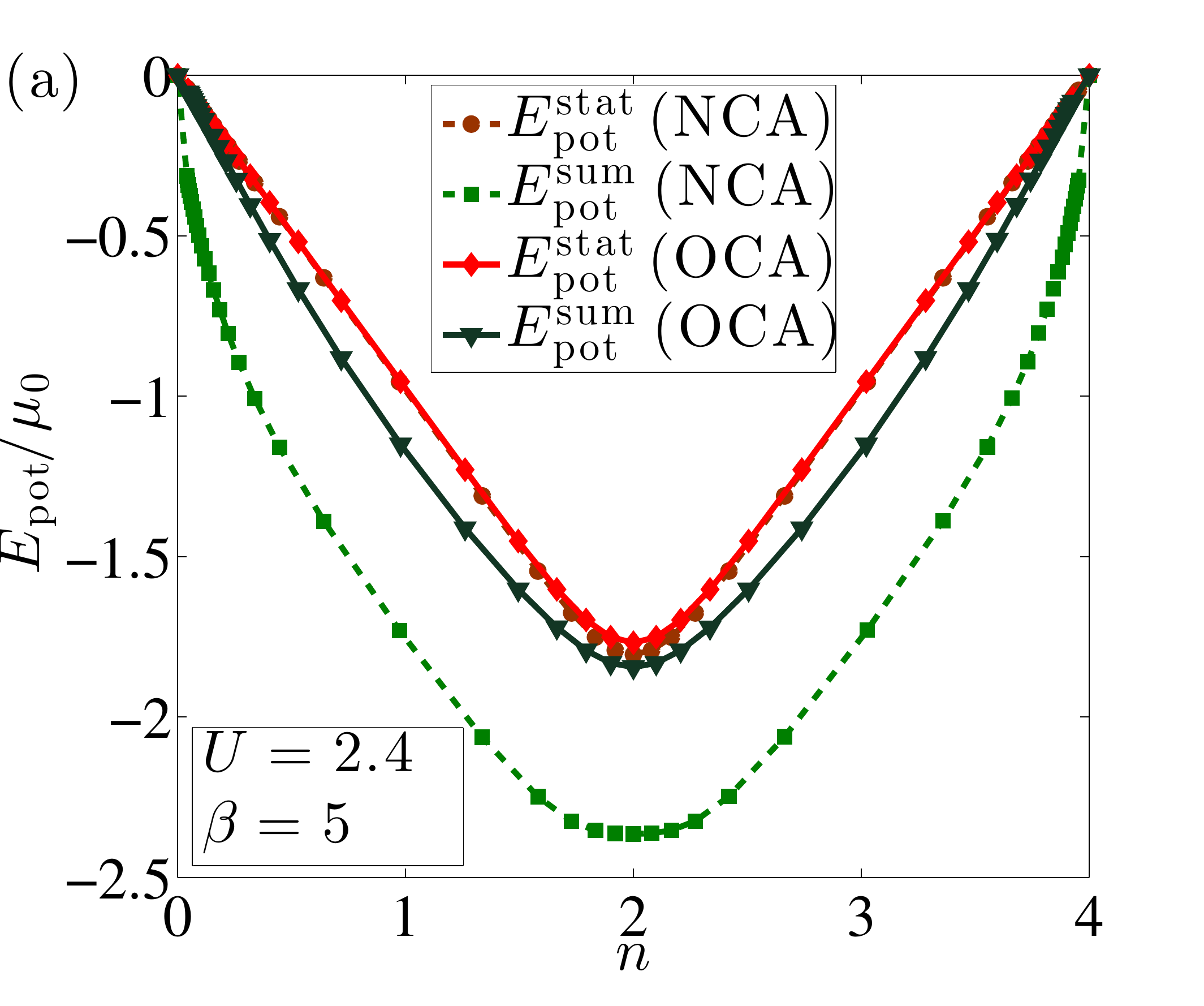}
\includegraphics[width=0.8\linewidth]{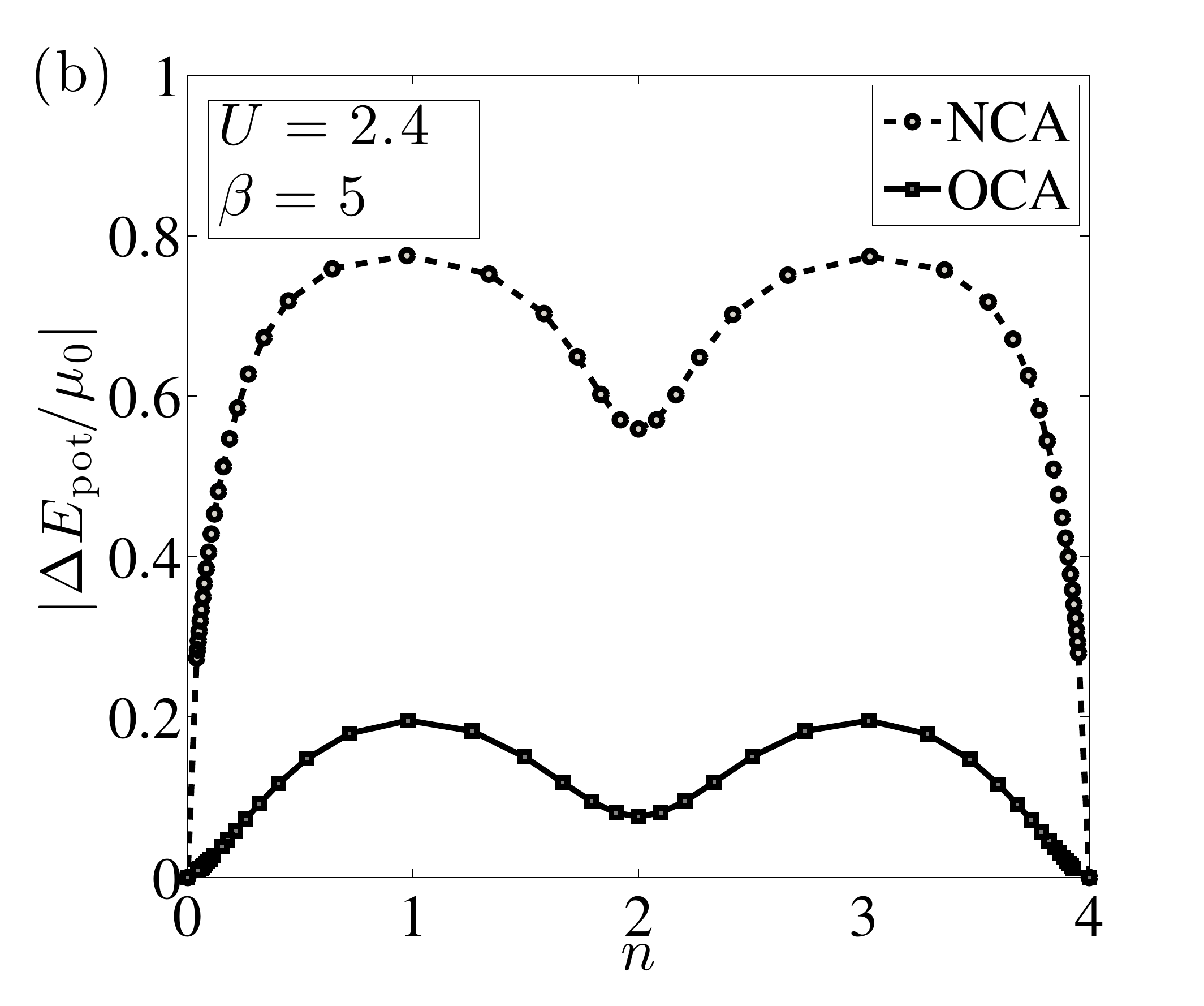}
\caption{(Color online) (a) The potential energy obtained from the two expressions Eqs.~\eqref{eq:Epot(1)} and \eqref{eq:Epot(2)} as a function of the impurity electron density $n$. (b) The ratio $|\Delta E_{\rm pot}/\mu_0|$ quantifying the sum rule violation for the same data as in (a).}
\label{fig:Epot}
\end{figure}
We now preset numerical results for the potential energy sum rule Eq.~\eqref{eq:Epot(2)}. Because we expect the biggest discrepancy for the spinful model, we restrict our discussion to this case. Figure~\ref{fig:Epot}(a) shows the (particle-hole symmetric) potential energy [Eqs.~\eqref{eq:Epot(1)} and \eqref{eq:Epot(2)}] as function of the total electron density on the impurity. We have normalized the curves with respect to  $\mu_0$, the chemical potential at half filling ($\mu_0=U$ for the spinful model). Note that the value for $E_{\rm pot}^{\rm sum}$, which is obtained from the Matsubara sum of $G(i\omega_n)\Sigma(i\omega_n)$, changes markedly between NCA and OCA while the thermodynamic expectation value of the potential energy $E_{\rm pot}^{\rm stat}$ changes only by a few percent. We therefore conclude that $E_{\rm pot}^{\rm stat}$ is more accurate, in accordance with the result of a perturbative expansion in the hybridization strength. 

Figure~\ref{fig:Epot}(b) shows the ratio $|\Delta E_{\rm pot}/\mu_0|$ for the same data as in (a). The sum rule violation is biggest around $n=1$ and $n=3$ but is smaller at half-filling and vanishes in the empty $(n=0$) or filled $(n=4$) limit.

\section{Benchmarking}
\label{sec:benchmark}
\begin{figure}
\includegraphics[width=0.75\linewidth]{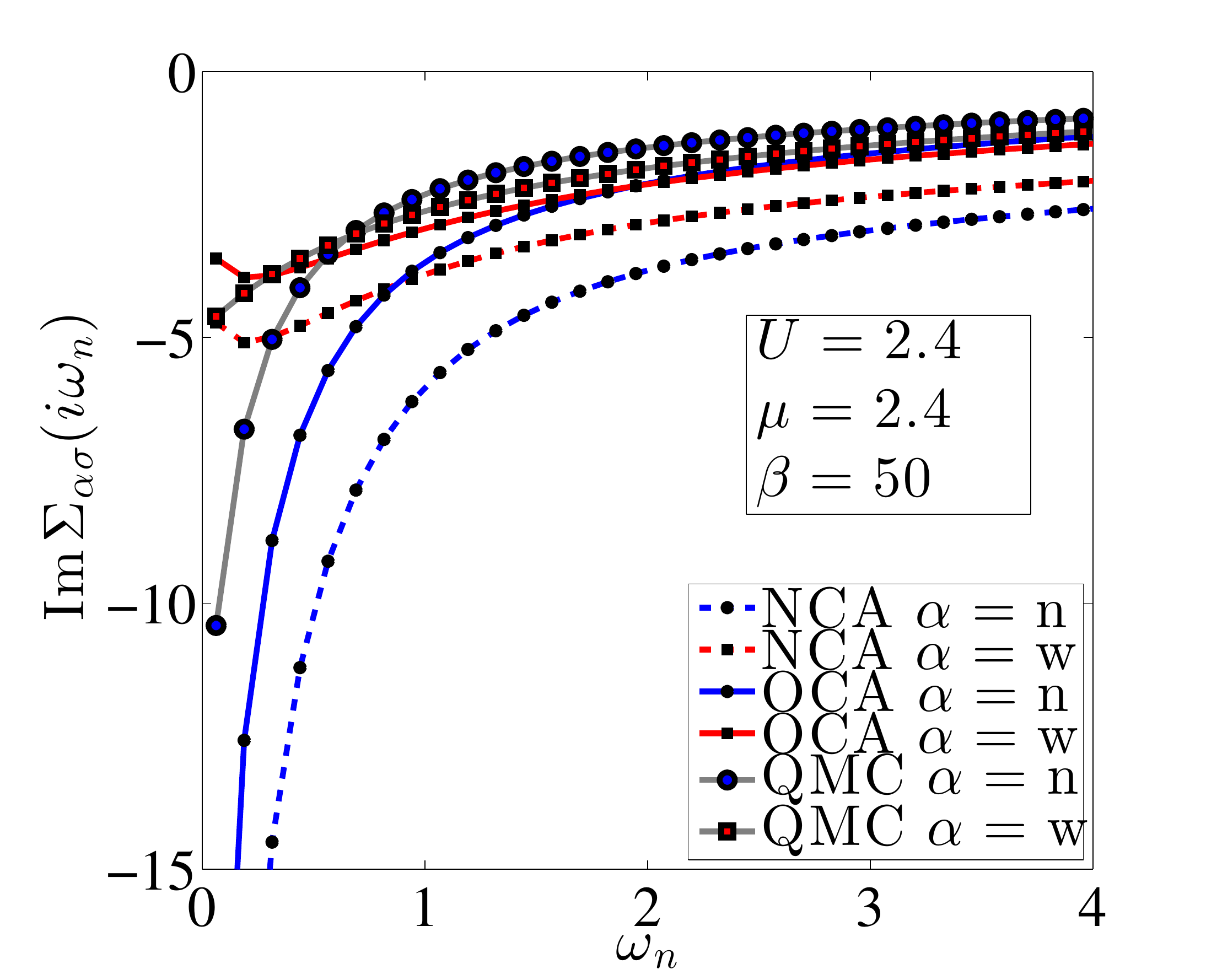}
\caption{(Color online) Low-frequency behavior of the imaginary part of the self-energy in the NCA and OCA for the spinful model compared with continuous time quantum Monte Carlo results [Ref.~\onlinecite{WangMillis:2010}] for the parameters specified in the plot. The sum rule violation for $\Sigma_1$ is shown in Fig.~\ref{fig:Sigma1_sf}.}
\label{fig:Sigma_sf}
\end{figure}

We now turn to a direct comparison of the NCA/OCA with continuous time quantum Monte Carlo (CT-QMC) data\cite{WangMillis:2010} for the two-orbital quantum dot model introduced in Sec.~\ref{sec:QD}. This allows us to directly address the accuracy of the NCA/OCA self-energy. The calculations were performed at particle-hole symmetry where the sum rule for the Hartree shift is exact within NCA/OCA for the parameters of Figs.~\ref{fig:Sigma1_sl} and Figs.~\ref{fig:Sigma_sf}. We find that the degree to which the sum rule for $\Sigma_1$ is violated gives a good estimate of the overall accuracy of the approximate self-energy.
\begin{figure}
\includegraphics[width=0.85\linewidth]{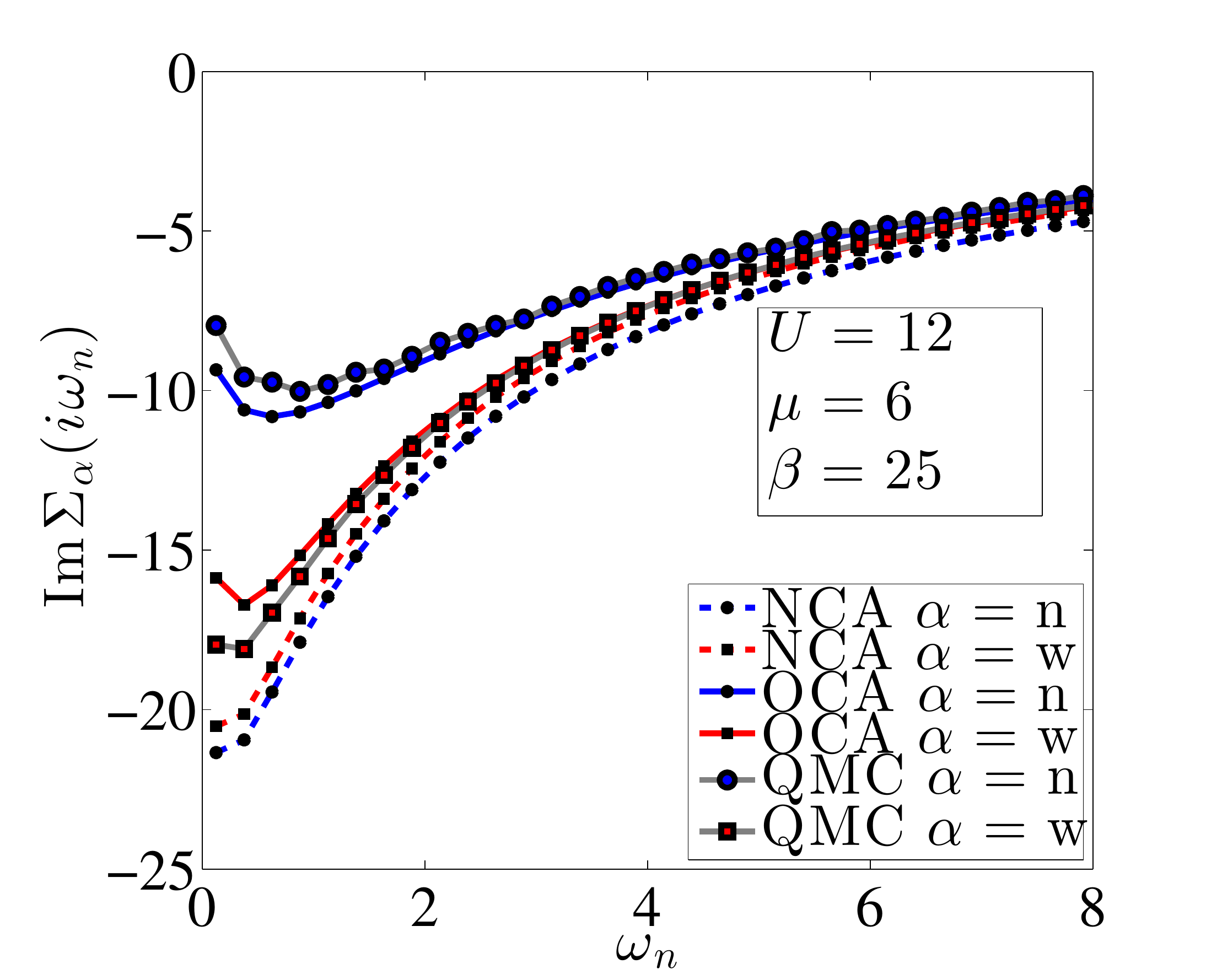}
\caption{(Color online) Low-frequency behavior of the imaginary part of the self-energy in the NCA and OCA for the spinless model compared with continuous time quantum Monte Carlo results [Ref.~\onlinecite{WangMillis:2010}] for the parameters specified in the plot. The sum rule violation for $\Sigma_1$ is shown in Fig.~\ref{fig:Sigma1_sl}.}
\label{fig:Sigma_sl}
\end{figure}

Figure~\ref{fig:Sigma_sl} shows the imaginary part of the self-energy as function of $\omega_n$ for the spinless model at particle-hole symmetry for an interaction $U=12$. For this large value of the interaction, the NCA prediction for the self-energy of the wide orbital is relatively close to the exact result. However, the NCA overestimates the self-energy for the narrow orbital by about a factor of two. The NCA thus fails to even qualitatively reproduce the subtle distinction between the two inequivalent orbitals arising from the orbital asymmetry of the hybridization. The inclusion of the one-crossing approximation substantially improves the results. These observations are in agreement with the results for the sum rule violation of the coefficient $\Sigma_1$ presented in Fig.~\ref{fig:Sigma1_sl}.

For the spinful model, we find that the agreement is less quantitative. Figure~\ref{fig:Sigma_sf} shows the imaginary part of the impurity-self energy at particle-hole symmetry for an inter-orbital interaction $U=2.4$. As compared to the exact result, both NCA and OCA predict a more insulting behavior for the narrow orbital. By extrapolating $\omega_n{\rm Im}\,\Sigma_{\rm n}(i\omega_n)$ to $\omega_n\rightarrow 0$ we find that OCA gives a gap for the narrow level which is almost twice the value found in CT-QMC. On the other hand, the self-energy for the broader level has metallic characteristics in the NCA/OCA while it is weakly insulating in the CT-QMC. The cause for these errors can be attributed to the fact the metal-insulator transition line is inaccurately predicted by NCA/OCA. Again, the overall accuracy is consistent with the degree the sum rule for $\Sigma_1$ is violated, as shown in Fig.~\ref{fig:Sigma1_sf}.
\section{Conclusions}
\label{sec:conclusions}
In summary, we reviewed the self-consistent hybridization expansions for multi-orbital quantum impurity models in the NCA and the OCA. We tested the degree to which these approximations respect several sum rules which hold in the exact theory. We focused on three examples. The first two were obtained from the analysis of the high-frequency expansion of the impurity self-energy. We found that already the static contribution (Hartree shift) can not be obtained from the thermodynamic expectation which fixes it in the exact theory. Similarly, the sum rule for the coefficient of the $1/\omega_n$ term is violated. The third example which we have studied is a sum rule which relates the potential energy to the Matsubara sum of $G(i\omega_n)\Sigma(i\omega_n)$. 

We note here that the observed sum rule violations are not incompatible with the fact that the NCA and OCA are $\Phi$-derivable conserving approximations. $\Phi$-derivability  ensures that various equivalent representations of the partition function based on integration of thermodynamic (static) quantities remain equivalent in the approximate treatment. However, $\Phi$-derivability does not ensure that sum rules, which relate dynamic to static properties, are satisfied.

In addition to the investigation of the above mentioned sums rules, we also benchmarked the NCA/OCA against exact CT-QMC results for a two-level quantum dot model with asymmetric coupling to two leads. From these different tests, we conclude that the NCA/OCA performs less satisfactory for weak interactions, away from particle-hole symmetry and in situations with multiple (potentially inequivalent) orbitals. In situations where exact results are not available, all three test, i.e.~the Hartree shift and the $1/\omega$ term of the self-energy as well as the potential energy sum rule, all provide simple tools to estimate the quality of the NCA/OCA. In our experience, the $1/\omega$ term is particularly informative and we suggest to use its relative error as a rule of thumb to address the accuracy of the approximation.

The error in the Hartree approximation has implications for the use of the NCA or OCA as impurity solvers for dynamical mean field theory.  A crucial aspect of the ``DFT+DMFT'' method \cite{Georges:2004,Held:2006,Kotliar:2006,Amadon:2008} which adds correlations to band theory is the ``double counting correction'' which is introduced to correctly place the energy of the correlated level relative to other orbitals in the material \cite{Anisimov:1991,Amadon:2008,Karolak:2010,Wang:2012}. The Hartree shift enters the computation of the double counting correction in an essential way, and if it is not reliably estimated then the physics is likely not to be correctly represented.

Both NCA and OCA are designed for the strong correlation limit and their accuracy is essentially controlled by the ratio of the hybridization to the local interaction. For the models we studied, we found that the NCA gives poor results for the self-energy even if the interaction is large compared to the hybridization. For example, the NCA self-energy does not reproduce the orbital asymmetry in a qualitative way. On the other hand, we found that the OCA clearly improves over the NCA
giving in particular a much improved account of the orbital asymmetries and a much smaller error in the sum rules. However, for moderate correlations it wrongly locates the transition point at which a gap opens in the spectrum and this can lead to qualitative errors in the low frequency portions of the spectrum. The probable magnitude of these errors can be estimated from the errors in the sum rule relating the coefficient of $1/\omega$ in the self energy to an expectation value. If this sum rule is reasonably well ($\lesssim 15\%$) obeyed, the small computational cost (relative to quantum Monte Carlo) of the OCA makes this an attractive choice for study  of the strongly interacting limit in a semi-quantitative way. 

An interesting application of the NCA/OCA involves nonequilibrium studies such as interaction quenches or switching on of an electric field.\cite{Eckstein:2010,Eckstein:2010b,Gull:2011b,Werner:2012} In these nonequilibrium systems, the imaginary time expansion is replaced by a unitary propagation on the Keldysh contour. By comparison with exact CT-QMC, it was found that the NCA/OCA works reasonably well for short enough time scales (as compared to the inverse of the hybridization strength).\cite{Gull:2011b} Other quantities, e.g. the relaxation to the steady state in the long-time limit, are markedly different from QMC results. We surmise that similar internal consistency checks exist also for real-time propagation which may be used for assessing the quality of nonequilibrium simulations.

\acknowledgements
We thank Xin (Sunny) Wang for providing us with the CT-QMC data of Ref.~\onlinecite{WangMillis:2010}. AR and GAF gratefully acknowledge financial support through ARO Grant No. W911NF-09-1-0527, NSF Grant No. DMR-0955778, and by grant W911NF-12-1-0573 from the Army Research Office with funding from the DARPA OLE Program. AR was partially supported by the Swiss National Science Foundation. EG and AJM acknowledge financial support through NSF-DMR-1006282. Part of the numerical calculations were performed at the Max Plank Institute for the Physics of Complex Systems in Dresden.

\bibliography{biblio}

\end{document}